\documentclass[12pt]{article}
\usepackage{epsfig}
\usepackage{amsmath,amssymb,amsthm,amscd}

\def\d{{\rm d}}

\def\2{{1\over2}}
\let\<=\langle \let\>=\rangle
\def\new#1\endnew{{\bf #1}}
\def\ifundefined#1{\expandafter\ifx\csname#1\endcsname\relax}
\ifundefined{draftmode}\else    \input draftmode        \fi
\let\Msize=\footnotesize             
\def\BM{\Msize\begin{matrix}}           \def\EM{\end{matrix}}
\def\MN M:#1 #2 N:#3 #4 {{(#1_{#2},#3_{#4})}}
\def\MNH M:#1 #2 N:#3 #4 H:#5,#6 [#7]{{(#1_{#2},#3_{#4})^{#5,#6}_{#7}}}

\newcommand{\ds}{\displaystyle}

\def\dd{\mathrm{d}}

\newcommand{\half}{{1\over 2}}

\newcommand{\be}{\begin{equation}}
\newcommand{\ee}{\end{equation}}
\newcommand{\bea}{\begin{eqnarray}}
\newcommand{\eea}{\end{eqnarray}}

\def\half{\frac{1}{2}}

\begin{document}
\begin{titlepage}
{}~ \hfill\vbox{ \hbox{} }\break

\rightline{Bonn-TH-09-01}
\rightline{CERN-PH-TH/2008-236} \vskip 1cm

\centerline{\Large \bf Holomorphicity and Modularity in}  \vskip 0.5 cm
\centerline{\Large \bf Seiberg-Witten Theories with Matter}  \vskip 0.5 cm
\renewcommand{\thefootnote}{\fnsymbol{footnote}}
\vskip 30pt \centerline{ {\large \rm Min-xin Huang
\footnote{minxin.huang@cern.ch} and Albrecht Klemm
\footnote{aklemm@th.physik.uni-bonn.de} }} \vskip .5cm \vskip 30pt
\vspace*{4.0ex}
\begin{center}
{\em $^\star$ Theory Division, Department of Physics, CERN,  CH-1211 Geneva, Switzerland}\\ [3 mm]
{\em $^\dagger$\ \ Bethe Center for Theoretical Physics and }\\ [1mm] 
{\em Physikalisches Institut, Universit\"at Bonn, D-53115 Bonn, FRG}
\end{center}

\setcounter{footnote}{0}
\renewcommand{\thefootnote}{\arabic{footnote}}
\vskip 100pt
\begin{abstract}
We calculate the gravitational corrections to the effective action of N=2
SU(2) Seiberg-Witten theory with matter using modularity, the holomorphic 
anomaly equation and expected behavior at the boundaries of the moduli 
space. As in pure gauge theory we show that the gap condition at the dyon 
singularities completely fixes the gravitational corrections. We discuss the 
behavior of the gravitational corrections at the conformal points. We 
compare our results with the recursive solution of the loop equation in the 
matrix model approach, which provides in addition open amplitudes.
\end{abstract}
\end{titlepage}
\vfill \eject


\newpage

\baselineskip=16pt

\tableofcontents

\section{Introduction}
\label{intro}
It was demonstrated in~\cite{Huang} for pure $N=2$ supersymmetric 
$SU(2)$ gauge theory that the modular symmetry, the holomorphic 
anomaly equation and the gap condition at the singularities, where 
dyons become massless, fix all gravitational corrections. 
Here we will  extend this approach to asymptotically free 
$N=2$ gauge theories with matter. The method we provide extends to 
several integrable systems which are connected to $N=2$ gauge theories. Let us first review 
these interrelations. 

When the gauge theory is coupled to gravity, the coupling of the selfdual 
curvature tensor $R_+$ to the graviphoton field strength $F_+$, i.e.  
$F^{(g)}({\underline a}) R_+^2 F_+^{2g-2}$, depends on the moduli fields
${\underline{a}}$ and this dependence is exactly calculable within the 
topological sector of the theory. An important point is that in different 
regions of the moduli there are different canonically conjugated coordinates 
and the different expansion of the $F^{(g)}$ are not 
just analytic continuations of each other. They are rather related by the 
wave function transformation of $Z=\exp(\sum_{g=0}^\infty\lambda^{2g-2} 
F_g({\underline a}))$~\cite{wittenwavefunction}, which is closely 
related to the modular and (an)holomorphic properties of the 
$F^{(g)}({\underline a})$~\cite{Aganagic:2006wq,Huang}. E.g. the weak coupling regime 
the $F^{(g)}({\underline a})$ can be calculated from the field theory point 
of view using localization in the space-time instanton moduli space~\cite{Nekrasov1,Nekrasov2}, 
but to obtain from this expansion the dual expansions e.g. at the monopole points or the
conformal points, also known as Argyres-Douglas points, one has to use the
wave function transformation~\cite{wittenwavefunction,Aganagic:2006wq,Huang}.

$N=2$ gauge theories are related to Type II string 
theory on non-compact Calabi-Yau manifolds by geometric 
engineering~\cite{Katz:1996fh}. It has been checked that 
the holomorphic gauge coupling functions and the BPS 
masses of the gauge theory~\cite{SW1,SW2} can also be obtained 
from the topological sector of the Type II string 
theory on this backgrounds in a double scaling limit which 
decouples $\alpha'$- and Planck scale
effects~\cite{Kachru:1995fv,Katz:1996fh}. 
In particular~\cite{Katz:1996fh} discusses the geometric engineering 
of non-compact toric Calabi-Yau spaces for $SU(N)$ gauge theories with few 
fundamental matter fields.  In the large radius region
of the topological string theory the $F^{(g)}({\underline t})$ are higher
genus world-sheet amplitudes, which depend on the K\"ahler moduli ${\underline t}$. 
The field theory $F^{(g)}(a)$ can be obtained in the limit mentioned  above~\cite{KMT}. 
A third approach is to calculate the $F^{(g)}(a)$ in the matrix model formalism suggested 
by~\cite{DV1,DV2} in an $\frac{1}{N}$ expansion~\cite{KMT,Dijkgraaf:2002yn}.

The common mathematical structure of $N=2$ Seiberg-Witten theories, 
topological string on non-compact Calabi-Yau spaces and matrix models 
is a family of Riemann surface ${\cal C}$ equipped with a meromorphic 
differential $\lambda$, which has in general non-vanishing residua. 

In the Seiberg-Witten approach to supersymmetric gauge theories ${\cal C}$ is the 
Seiberg-Witten curve and $\lambda$ the Seiberg-Witten differential~\cite{SW1,SW2}.
$({\cal C},\lambda)$ can be re-derived from the space-time instanton calculus as 
was demonstrated in~\cite{Nekrasov2}\footnote{Using~\cite{Eynard:2008mt} 
one can construct the space-time instanton sums of~\cite{Nekrasov2} a 
direct link to the matrix model~\cite{Klemm:2008yu}.}. In topological 
string theory on non-compact Calabi-Yau spaces the data $({\cal C},\lambda)$ arise 
via mirror symmetry in the B-model geometry~\cite{Katz:1996fh,Hori:2000kt}.
In the matrix model approach ${\cal C}$ is the spectral curve and 
$\lambda$ the differential defining the filling 
fractions and the one point function~\cite{DV1,DV2,Nekrasov2,Marino:2006hs,Klemm:2008yu}. The calculations 
of~\cite{Nekrasov1,Nekrasov2} using ST-instanton calculus, 
of~\cite{Huang,Haghighat:2008gw} within the topological 
B-model, and of~\cite{EO,Marino:2006hs} utilizing the matrix model 
recursions suggest that the higher genus information can be 
completely reconstructed from $({\cal C},\lambda)$. 

The $B$ model approach is particularly efficient. It uses for the reconstruction 
the recursion relation in the genus, known as holomorphic anomaly equation~\cite{BCOV}. 
The latter has a kernel, the holomorphic ambiguity, which is strongly 
constrained by the symmetry group of theory, but certain discrete data 
have to be fixed by additional arguments. For pure gauge theory the gap 
condition at the conifold fixes these discrete data completely~\cite{Huang}.   
  
In this work we consider the asymptotically free $SU(2)$ with  $N_f=1,2,3$ 
hypermultiplets in  the fundamental representation.  
Similar as the topological string theory on the 
canonical line bundle over a del Pezzo surface 
with more than one K\"ahler parameter~\cite{Haghighat:2008gw}, 
gauge theories with matter have more parameters 
than the complex moduli space of the Riemann surface. In gauge theory 
these parameters are simply the masses of the hypermultiplets. While they make
the form of the holomorphic ambiguity more 
complicated, the requirement that the gap  exists for all values 
of the additional parameter imposes stronger conditions. 
We find that the latter overconstrain  the system and that an unique 
solution exists. The gravitational corrections for gauge theories 
with matter are therefore completely integrable. 

One advantage of the method~\cite{Huang} is that it 
provides the $F^{(g)}$ in all  regions in the moduli space and not 
just in the asymptotically  free region. Argyres, Plesser, 
Seiberg and Witten found particularly interesting points in 
the moduli space of $SU(2)$ gauge theory with fundamental matter, where different 
conformal theories arise~\cite{Argyres}. Using the formalism~\cite{Huang}, 
one can analyze the full topological sector of these conformal theories. 
The coordinate choice and the structure of the topological gauge theory 
amplitudes near conformal points is very similar as for topological 
string theory near orbifold points~\cite{Bouchard:2007ys,Bouchard:2008gu}.

The paper is organized as follows: In Section~\ref{sectionmassless}, 
we first consider the simple case of theory with massless matter. 
Here the Coulomb modulus $u$ is related  to the modular parameter 
$\tau$ by $J(\tau)=R_{N_f}(u)$, where $R_{N_f}(u)$ is
a rational function and $J$ is the modular invariant $J$-function.
The Coulomb moduli space is then a ramified finite multicover of the quotient 
of the upper half plane ${\rm Im}(\tau) >0$ by ${\rm PSL}(2,\mathbb{Z})$ 
and $u(\tau)$ has very simple transformation properties 
w.r.t. ${\rm PSL}(2,\mathbb{Z})$. We review the direct integration, 
which is an efficient way to solve the holomorphic anomaly equation 
and write the $F^{(g)}$ as polynomials of modular functions. In
Section \ref{sectionmassive} we generalize the analysis to the
massive $SU(2)$ Seiberg-Witten theory. We provide an algorithm
that is valid for the case of fundamental matter with arbitrary
mass, and study the $N_f=1$ case in details. This procedure can be
straightforwardly generalized to $N_f=2,3$. In Section
\ref{sectionsuperconfomal} we discuss some cases of special
fundamental hypermultiplet masses where two mutually non-local
singular points in the $u$ plane collide and a non-trivial
superconformal field theory appears. Section~\ref{locallimit} 
discusses the emergence of the gauge theory in the non-compact 
limit of a Calabi-Yau compactification. This provides a connection 
to the direct integration formalism developed in~\cite{Yamaguchi,
Grimm:2007tm,Alim:2007qj,Hosono}, which is useful to solve the 
gravitational couplings of higher rank gauge theories entirely 
from the solutions of the Picard-Fuchs equation. In Section
\ref{matrixsection-05-19} we turn to the matrix model approach. 
We find it particularly to solve the the $N_f=2$ theory by the 
approach of~\cite{EO}, point out some restrictions in the  
application of the open holomophic anomaly of~\cite{EMO} and 
discuss analytic properties of the open amplitudes.

\section{$SU(2)$ Seiberg-Witten theory with massless
hypermultiplets} \label{sectionmassless}

The Seiberg-Witten curves\footnote{For the $N_f=0$ we use the family of curves
  of~\cite{Klemm:1994qs} rather than the one of~\cite{SW1}. This distinction
  plays a r\^ole in  establishing the matrix model connection in section
  \ref{matrixsection-05-19}.}~\cite{SW1, SW2}  
${\cal  C}_1$ for $\mathcal{N}=2$ supersymmetric $SU(2)$ gauge 
theory with $N_f<4$ flavors are  families of elliptic
curves given by~\footnote{One can do a change of variable and write the
Seiberg-Witten curve in all cases as
\begin{equation} 
y^2=(x^2-u)^2-\Lambda^{2N_c-N_f}\prod_{i=1}^{N_f}(x+m_i)
\end{equation}
\cite{D'Hoker}. Here we use the original notations in \cite{SW2}.}
\begin{equation} 
\label{curves}
y^2=C(x)^2-G(x)\ ,
\end{equation} 
where $C(x)$ and $G(x)$ are defined as
\begin{equation}
\begin{array}{rlrlrl} 
N_f\!\!\!\!&=0:\! & C(x)\!\!\!&=x^2-u,
& G(x)\!\!\!\!&=\Lambda^4, \\[2 mm]
N_f\!\!\!\!&=1:\! & C(x)\!\!\!&=x^2-u,
& G(x)\!\!\!&=\Lambda^3(x+m_1),  \\[2 mm]
N_f\!\!\!\!&=2:\! & C(x)\!\!\!&=x^2-u+\frac{\Lambda^2}{8},
& G(x)\!\!\!&=\Lambda^2(x+m_1)(x+m_2), \\[2mm]
N_f\!\!\!\!&=3:\! & C(x)\!\!\!&=x^2-u+\frac{\Lambda}{4}(x+\frac{m_1+m_2+m_3}{2}),
& G(x)\!\!\!&=\Lambda(x+m_1)(x+m_2)(x+m_3)\ . \\[2mm] 
\end{array} \nonumber
\end{equation}
Here $u$ is the modulus parametrizing the Coulomb branch 
and $m_i$ are the masses of the hypermultiplets. 
The genus one curves ${\cal  C}_1$ have two periods $a_D$ and
$a$ of the meromorphic Seiberg-Witten differential $\lambda$ over 
the $b$ and the $a$ cycle in $H^1({\cal C}_1,\mathbb{Z})$ respectively. 
The meromorphic one form $\lambda$ can be written with $'=\frac{d}{dx}$ as 
\begin{equation} 
\lambda=\frac{\sqrt{2}}{4 \pi i} \frac{x dx }{y}\left[\frac{C(x) G'(x)}{2
    G(x)}- C'(x)\right] \ .
\label{lambda}
\end{equation}
Physically the periods are the vev's of the scalar component 
of the $\mathcal{N}=2$ vector multiplet containing the photon and its dual 
in the infrared respectively. 

We are interested in calculating the instanton expansion of the 
prepotential $F^{(0)}$ and its higher genus generalization 
$F^{(g)}$ all over the moduli space. In the asymptotically free region 
the prepotential $F^{(0)}(a)$ for the electric $U(1)$ theory, which  
determines the exact gauge coupling of the $N=2$ Super-Yang-Mills theory, 
is related to the periods $a_D$ and $a$ by rigid special geometry
\begin{eqnarray}
\label{F0}
\frac{\partial F^{(0)}}{\partial a}=a_D\ .
\end{eqnarray}
As pointed out in \cite{SW1} in the dual magnetic region where 
$a_D$ is small, the theory is more suitably described by a magnetic $U(1)$, 
whose prepotential is given by $\frac{\partial F_D^{(0)}}{\partial a_D}=a$. 

The higher genus terms $F^{(g)}(a)$ describe the exact moduli dependence 
of the gravitational corrections
\begin{equation} 
\label{gravitationalcorrections}
F^{(g)} F_{+}^{2g-2}R_{+}^2\ .
\end{equation} 
 in the effective Lagrangian, which  encodes the coupling of 
the gauge theory to $N=2$ supergravity.  Here $F_{+}$ and $R_{+}$ are the 
self-dual part of the graviphoton field strength and of the Ricci 
curvature respectively. In the weak coupling region 
$u\rightarrow \infty$ we can compare  $F^{(g)}(a)$ to the 
localization calculations in \cite{Nekrasov1}. The relation of the dual 
$F_D^{(g)}(a_D)$ to the $F^{(g)}(a)$, follows from the quantum 
mechanical wave function transformation of 
$Z=\exp(\sum_{g=0}^\infty \lambda^{2g-2} F^{(g)})$  and was studied in 
this context in~\cite{Aganagic:2006wq}.

In this section, we first consider the simple case where all the
hypermultiplets are massless. In this case $u(\tau)$ is a 
modular invariant function under the projective action on 
the gauge coupling $\tau\propto \frac{\partial^2}{\partial^2 \tau} F^{(0)}$ of
the group $\Gamma_{N_f}\in {\rm  PSL}(2,\mathbb{Z})$~\cite{Nahm}.  To find the rational function $R_{N_f}(u)$ discussed in 
section~\ref{intro} one simply brings the curve (\ref{curves}) 
into Weierstrass form
\begin{equation} 
\label{weierstrassform}
y^2=4x^3-g_2(u)x-g_3(u)\ .
\end{equation}
The rational function is then determined by
\begin{equation} 
\label{J}
J(\tau)=\frac{E_4(\tau)^3}{E_4^3(\tau)-E_6(\tau)^2}=\frac{g_2(u)^3}{g_2(u)^3-27g_3(u)^2}=:R_{N_f}(u)\ .
\end{equation}
Here  
\begin{equation}
\Delta = g_2(u)^3-27g_3(u)^2\ 
\label{Delta}
\end{equation}  
is the discriminant of the curve. The group $\Gamma_{N_f}$ is the quotient 
of ${\rm  PSL}(2,\mathbb{Z})$ by the group
interchanging the roots of $u(J)$ and has been determined in~\cite{Nahm}. In the 
mathematical literature the invariant $u(\tau)$ is sometimes 
called the `Hauptmodul' of $\Gamma_{N_f}$.
    
The Picard-Fuchs differential equations are fulfilled by 
all periods over the cycles of ${\cal C}_1$. In the massless case 
$\lambda$ has no non-vanishing residua and the two periods $a_D$ and $a$
fulfill a second order Picard Fuchs differential equations\footnote{In the massive case 
discussed in section~\ref{sectionmassive} there is also a cycle 
encircling the pole of $\lambda$ picking up the 
residuum and the differential equations are third order.}, which were derived for $N_f=1,2,3$ in~\cite{Ito}
\begin{equation} \label{PFmassless}
p(u)\frac{d^2\Pi}{du^2}+\Pi=0\ ,
\end{equation}
where
\begin{equation}
\begin{array}{rlrlrlrl}  
N_f&=1:  &  p(u)&= 4u^2+\frac{27\Lambda_1^6}{64u} & N_f&=2:  &  p(u)&= 4(u^2-\frac{\Lambda_2^4}{64})  \\[2 mm]
N_f&=3:  &  p(u)&= u(4u-\frac{\Lambda_3^2}{64})\  .  &    &     &      &  \\ [2 mm]
\end{array} 
\end{equation} 
Using the leading behavior of $F^{(0)}$ and  $F_D^{(0)}$ from the 1-loop $\beta$-function
and analytic continuation we find $\Pi=\left(\begin{array}{c} a\\ a_D \end{array}\right)$ 
as linear combination of the solutions to (\ref{PFmassless}).
We will set the dynamical scales $\Lambda_1=2^{\frac{2}{3}}$,
$\Lambda_2=2$, $\Lambda_3=4$ in order to match the convention in
the instanton counting calculations in \cite{Nekrasov1,
Nekrasov2}.

While in the pure $SU(2)$ gauge there is a $Z_2$ symmetry 
acting on the $u$ plane, the discrete symmetries of the $u$ 
plane in $N_f=1,2,3$ cases are $Z_3$
symmetry, $Z_2$ symmetry and no symmetry respectively \cite{SW1,
SW2}. As we will see, these different discrete symmetries acting
on the $u$ plane in the three cases $N_f=1,2,3$ play significant
role in determining the qualitative features of the solutions. We
will find the structure of $N_f=2$ solution closely resembles that
of the case of pure gauge theory in \cite{Huang}, while the cases
$N_f=1,3$ have some different qualitative features respectively.

In the next section we will review the direct integration approach 
for solving $F^{(g)}$. Thereafter we discuss the $N_f=1,2,3$ cases 
one by one.

\subsection{Topological string amplitude $F^{(g)}$
as a polynomial of $\hat{E}_2$} \label{sectionmasslessdirect}

The main goal is to solve the topological sector of the theory 
and give in particular the $F^{g}(u)$ everywhere in the Coulomb 
moduli space. To this end we first extend the direct integration 
method of the holomorphic anomaly equations to the $SU(2)$ gauge 
theory in this section to the case with massless flavors and 
in section~\ref{sectionmassive} to the case with massive 
flavors. This approach was applied to $N_f=0$ 
in~\cite{Huang} and solved the theory completely using the gap 
condition. The point is to show this for theories with 
flavors as well.

The holomorphic anomaly equations of \cite{BCOV} read
\begin{equation}
\begin{array}{rl}
\partial_a\partial_{\bar{a}}F^{(1)}&=\frac{1}{2}C_{aaa}C_{\bar{a}}^{aa},\\[2mm] 
\bar \partial_{\bar a} F^{(g)}&=\frac{1}{2} C_{\bar a}^{aa} \left(D_a D_a F^{g-1} +
\sum_{g=1}^{g-1} D_a F^{(g-h)} F^{(h)}\right),\qquad {\rm for}\ g>1\ . 
\end{array}
\label{anomaly}
\end{equation}    
Here we used the coordinate $a$ introduced in the last section, but 
the equations are of course covariant. We further introduced the Yukawa 
triple coupling and the connection $D_a$, whose calculation from the solutions 
of the Picard Fuchs equation are discussed below and more generally in 
section (\ref{specialgeometry}). First $F^{(0)}(a)$ follows from the 
solution $\Pi$ to (\ref{PFmassless}) via (\ref{F0}) up to an irrelevant 
constant. We define then the three point Yukawa coupling as
\begin{equation}
C_{aaa}=\frac{\partial^3F^{(0)}}{\partial a^3}= -2\pi im_0
\frac{d\tau}{d a}=:\xi
\label{yuk1}
\end{equation}
Our normalization convention is $m_0=2, 1,1, \frac{1}{2}$ for the 
cases of $N_f=0,1,2,3$. Mathematically $\tau$ is the modular parameter and
physically 
\begin{equation}
\tau= -\frac{1}{2\pi i m_0}\frac{\partial^2F^{(0)}}{\partial a^2}
\label{tau}
\end{equation}
a combination\footnote{Here we used the normalization of~\cite{SW2}.} 
of the gauge coupling and the theta angle $\tau=\frac{\theta}{\pi} + i \frac{8 \pi}{g^2}$. 
Note that  $\tau_2={\rm Im}(\tau)=\frac{\tau-\bar \tau}{2i}$ 
multiplies the kinetic term of the vector multiplet\footnote{The key requirement that the 
latter has to be positive suggested the occurrence of Riemann surfaces 
in this context, where $\tau_2$ is manifestly positive.}.
With the methods described in  Chapter 5 of~\cite{zagier} one can prove the following 
modular expression for $\xi$
\begin{equation} \label{xi}
\begin{array}{llll} 
N_f=0:& \qquad \xi=\frac{8\theta_2^2}{\theta_3^4\theta_4^4}&, 
\qquad N_f=1:&\qquad \xi=\frac{4\sqrt{6}E_4^{\frac{1}{2}}(E_4^{\frac{3}{2}}-E_6)^{\frac{1}{6}}}{(-1)^{\frac{1}{6}}(E_4^{\frac{3}{2}}+E_6)},\\
N_f=2:& \qquad \xi=\frac{4\theta_2^2}{\theta_3^4\theta_4^4}&, 
\qquad N_f=3:&\qquad \xi=\frac{8\theta_2^2}{\theta_3^4\theta_4^4},\\
\end{array}
\end{equation} 
where $E_k$ and $\theta_k$ are the standard Eisenstein series 
and Jacobi $\theta$ functions in the conventions~\cite{zagier}.      
Further the Weil-Petersson metric in the coordinate $a$ is given by
\begin{equation}
G_{a\bar{a}}=2\partial_a\partial_{\bar{a}}
\textrm{Re}(\bar{a}\partial_aF^{(0)})=4m_0\pi \tau_2 \ .
\label{metric}
\end{equation}
The connection $D_a$ comes entirely\footnote{In the global Calabi-Yau case
  there is an additional K\"ahler connection, as explained in
  section~\ref{specialgeometry}.}  
from the metric $G_{a\bar{a}}$, and is
\begin{equation} \label{connection}
\Gamma^{a}_{aa}=(G_{a\bar{a}})^{-1}\partial_a(G_{a\bar{a}})=-\frac{i}{2\tau_2}\frac{\partial
\tau}{\partial a}\ .
\end{equation}
Note that $\Gamma_{aa}^a$ vanishes in the holomorphic limit
$\bar{\tau}\rightarrow \infty$, confirming that the period $a$ is
a flat coordinate in this limit. 
 Finally we have 
\begin{equation}
C^{aa}_{\bar a}=G^{a\bar a} G^{a\bar a} \bar C_{\bar a\bar a\bar a} \ .
\label{yuk2}
\end{equation} 
In the following it will be important to keep track of the 
anti-holomorphic dependence $\Gamma_{aa}^a$ in (\ref{anomaly}) 
in order to recover the full $F^{(g)}(\tau,\bar{\tau})$ 
including its anti-holomorphic dependence. 

The holomorphic anomaly equation determines $F^{(g)}$ from 
lower genus data up to a holomorphic anomaly, which can be 
fixed by modularity and the gap condition. Let us start 
with genus one, which is somewhat special. It follows from 
(\ref{yuk1},\ref{metric},\ref{yuk2}) that the right hand side 
of the $g=1$ equation in (\ref{anomaly}) is $\frac{1}{8\tau_2^2}|
\frac{\partial \tau}{\partial a}|^2$. This can be integrated 
to
\begin{equation}
F^{(1)}=-\frac{1}{2} \log(\tau_2)-\log|\Phi(\tau)|^2\ ,
\end{equation}
where $\Phi(\tau)$ is modular form of weight $\frac{1}{2}$, 
which vanishes at the discriminant of the 
elliptic curve. In the simplest cases, e.g. 
$N_f=0$ one $\Phi(\tau)$ can be identified with 
the Dedekind $\eta$-function. Note that the transformation of $\Phi(\tau)$ as weight 
$\frac{1}{2}$ modular form cancels the transformation of the 
$-\frac{1}{2} \log(\tau_2)$ term. In general the 
form $\Phi(\tau)$ is determined by its modularity 
and its leading logarithmic behavior near 
$a_D^{(k)}=0$. It has been pointed out in~\cite{Vafa:1995ta} that 
$\log(a_D^{(k)})$ comes from the gravitational one-loop $\beta$ 
function and its prefactor is entirely determined by the 
massless spectrum at the critical point.

$F^{(1)}(\tau,\bar \tau)$ is an almost holomorphic modular 
function or form of weight 0\footnote{In slight abuse of notation we indicate 
almost holomorphic objects by writing a $\bar \tau$ dependence.}
and its $\tau$ derivative, which appears in (\ref{anomaly}), is an 
almost holomorphic form of weight 2. Modularity implies that this derivative 
contains the unique almost holomorphic modular weight two form
\begin{equation}
\label{E2hat}
\hat{E}_2(\tau,\bar \tau)=E_2(\tau)-\frac{3}{\pi \tau_2}\ ,
\end{equation}
where $E_2(\tau)$ is the holomorphic quasimodular 
second Eisenstein form~\cite{zagier}. Under modular transformations 
$\tau \mapsto \tau_\gamma=\frac{a\tau + b}{c \tau + d}$
with $\gamma\in \Gamma_1={\rm PSL}(2,\mathbb{Z})$ $E_2$ transforms
with an inhomogeneous shift
\begin{equation} 
E_2(\tau_\gamma)=(c \tau + d)^2 E_2(\tau)-\frac{ 6 i c}{\pi} (c \tau+d)\ .
\label{E2trans}
\end{equation}
This shift cancels the shift transformation of 
$-\frac{3}{\pi \tau_2}\ $, so that $\hat{E}_2$ transforms indeed as a honest  
weight two form. 

In calculating the right hand side of (\ref{anomaly}) one needs derivatives 
of modular forms of even positive weight. 
The covariant derivative $D$ in (\ref{anomaly}) written in terms of the 
$\tau$ coordinate is the so called Mass derivative 
\begin{equation} 
\hat{D}_\tau=\hat{\partial}_\tau-\frac{k}{4\pi\tau_2}\ . 
\end{equation}
Here $\hat \partial_{\tau}=\frac{1}{2 \pi i} \frac{d}{d\tau}$ and $k$ is 
the modular weight of the object acted on. The Mass derivative $\hat{D}_\tau$ 
has the important property that it maps almost holomorphic modular forms of weight $k$ into 
almost holomorphic modular forms of weight $k+2$. Modular invariance implies 
that each covariant derivative increases the leading power 
of  $\hat E_2$ by one and all powers of $\frac{1}{\tau_2}$
must combine with  $E_2(\tau)$ to form the shift invariant 
combination $\hat{E}_2(\tau,\bar \tau)$. From this follows the 
important fact that for $g\geq 2$ all anholomorphic dependence of $F^{(g)}$ is 
in $\hat E_2$ and we can replace
\begin{eqnarray}
\frac{d}{d\bar{\tau}}=\frac{d\hat{E_2}}
{d\bar{\tau}}\frac{d}{d\hat{E_2}}=\frac{3i}{2\pi\tau_2^2}\frac{d}{d\hat{E_2}}\ .
\end{eqnarray}
Furthermore we find that the anti-holomorphic
derivative in (\ref{anomaly}) combines with the
three point function as
\begin{equation}
\frac{2\partial_{\bar{a}}F^{(g)}}{C_{\bar{a}}^{aa}}=
24m_0\frac{dF^{(g)}}{d\hat{E_2}},~~~~ \textrm{for} ~g\geq 2\ .
\end{equation}
so that (\ref{anomaly}) can be written as 
\begin{equation} \label{Albrechtholomorphic1}
24m_0\frac{dF^{(g)}}{d\hat{E_2}} =
D_a^2F^{(g-1)}+\sum_{r=1}^{g-1}\partial_aF^{(r)}\partial_aF^{(g-r)}\ .
\end{equation}
Since the period $a$ is a quasimodular object of weight 1 and $\hat E_2$ is 
of weight $2$ one concludes that all $F^{(g)}$ have modular weight zero.
Combining the above facts it follows that the $F^{(g)}$ 
are inhomogeneous polynomials of degree $3(g-1)$ in 
$\hat E_2$ whose coefficients are holomorphic forms of 
negative weight so that  $F^{(g)}$ have weight zero.
Defining the following derivatives
\begin{eqnarray} \label{derivativeMass}
&&\partial_aF^{(r)}=-\frac{\xi}{m_0}\hat{\partial}
_{\tau}F^{(r)}=-\frac{\xi}{m_0}\hat{D}_{\tau}F^{(r)} \nonumber \\
&&
D_a^2F^{(r)}=(\partial_a-\Gamma_{aa}^a)\partial_aF^{(r)}=\frac{1}{m_0^2}\xi\hat{D}_\tau(\xi\hat{D}_{\tau}F^{(r)}
)\ , 
\end{eqnarray}
where we  used the connection (\ref{connection}) and the fact 
that $F^{(r)}$ and $\partial_aF^{(r)}$ have modular weight 
zero and $-1$ respectively, we  write the holomorphic anomaly equation as
\begin{eqnarray}
\label{BCOV-04-27} 24m_0^3\frac{dF^{(g)}}{d\hat{E_2}} =
\xi^2(\hat{D}_\tau^2F^{(g-1)}+\frac{\hat{D}_\tau\xi}{\xi}\hat{D}_\tau
F^{(g-1)} +\sum_{r=1}^{g-1}\hat{D}_\tau F^{(r)}\hat{D}_\tau
F^{(g-r)})\ .
\end{eqnarray}
This provides an unifying description of the Seiberg-Witten theory
with various number of massless flavors, depending only on $m_0$ and the 
holomorphic modular form of weight $-3$ given in (\ref{xi}).

In each integration step the coefficients of all nonzero powers 
of $\hat E_2$ are determined by (\ref{anomaly}), while an holomorphic ambiguity 
of modular weight zero can be added. Boundary conditions and modularity  
imply that this can be always written as $\xi^{2g -2}$ times a modular
form of weight $6(g-1)$. This reduces problem of fixing the ambiguity  
to the determination of a finite number of terms. The remaining task is solved 
by an analysis of the local form of the effective action, which we 
discuss next.   

A key concept in the analysis of effective action~\cite{SW2} is 
its transformation property under the modular group $\Gamma_{N_f}$ 
and the concept of local holomorphic coordinates in which the 
effective action is expanded near the critical points of the 
theory, where particles becoming massless~\cite{SW2}. 
In particular in the asymptotically free region of the 
gauge theory $a$ or more precisely $\frac{1}{a}$ is the 
correct small expansion parameter, while near points where 
a dyon of magnetic charge and electric 
charge $(q_m,q_e)$ in an $N=2$ hypermultiplet becomes 
massless, i.e. close to the components of the  discriminant 
locus of (\ref{curves}), $a_D^{(k)}=q^{(k)}_m a_{D}+q^{(k)}_e a$ 
is the small expansion parameter. In most cases, i.e. for 
$N_f=0,1,2$ the $Z_2$, $Z_3$ and $Z_2$ symmetry of the theories respectively 
relates the dyon points and the local expansions are the same, but for $N_f=3$ we find 
truly inequivalent dyon points.  

At the magnetic monopole point, also called conifold point, the 
leading term of the topological string amplitudes in the variable $a_{D}$, 
is determined by the $c=1$ string at the selfdual radius~\cite{Ghoshal:1995wm}. 
The $2g-1$ sub-leading terms are absent. This gap structure\footnote{Below we can always 
rescale $a_D$ so that $c=1$. The Bernoulli numbers $B_k$ are defined by  $\sum_{k=0}^\infty \frac{B_k x^k}{k!}=\frac{x}{e^x-1}$, 
i.e. $B_2=\frac{1}{6}$, $B_4=-\frac{1}{30}$, $B_6=\frac{1}{42}$ etc.} 
\begin{equation}  
F^{(g)\, D}= \frac{c^{g-1} B_{2g}}{2g (2g-2)a_D^{2g-2}}+{\cal O }(a_D^0) \ . 
\label{gapm}
\end{equation}  
has been observed in~\cite{Huang} and as explained in~\cite{HKQ} 
it originates indeed from integrating out massless particles in the 
Schwinger loop contribution to the higher derivative effective 
action. It was  used in~\cite{Huang} to fix the holomorphic ambiguity 
in the calculation of the gravitational 
corrections in pure $SU(2)$ Seiberg-Witten theory. Here we find 
that the gap occurs at all dyon points. Using the gaps and the 
leading coefficients we are able to fix in the  $N_f=0,1,2,3$ 
cases the holomorphic ambiguity genus for all genus, and obtain 
exact formulae for the gravitational corrections $F^{(g)}$ 
that sum up all instanton contributions at each genus $g$.

Let us finish this section with some comments on (\ref{BCOV-04-27}) 
and a calculation of the  leading terms in $\hat E_2$
First we note that the equations leading to (\ref{BCOV-04-27}) 
are invariant under the change $\hat{\partial}_\tau, E_2 \rightarrow 
\hat{D}_\tau, \hat{E}_2$, one may therefore as well take a ``holomorphic 
limit'' and replace $\hat{D}_\tau, \hat{E}_2$ with $\hat{\partial}_\tau, E_2$ in 
equation (\ref{BCOV-04-27}), without losing any information.
Furthermore the  holomorphic anomaly equation (\ref{BCOV-04-27}) provides a
very efficient way to compute topological string amplitudes. While in
the Feynman rule approach of BCOV the number of diagrams grows exponentially with
$g$, in the direct integration approach the number of terms in $F^{(g)}$ grows 
only with a power law with the genus $g$. This is similar as in the case of 
quintic Calabi-Yau three-fold studied in \cite{Yamaguchi, HKQ}.

The leading coefficients of $F^{(g)}$ as a polynomial of
$\hat{E}_2$ do not depend on the holomorphic ambiguity and can be
computed to very high orders. Suppose we denote the leading terms
by
\begin{eqnarray}
F^{(g)}=\frac{A^{(g)}}{(g-1)(1152m_0^3)^{g-1}}\xi^{2(g-1)}\hat{E}_2^{3(g-1)}+\cdots
\end{eqnarray}
where $\cdots$ denotes terms with lower powers of $\hat{E}_2$. One
can see that the coefficients $A^{(g-1)}$ do not depend on the
holomorphic ambiguities, since there is no $E_2$ in the
holomorphic ambiguities in all the models we study. Using the
holomorphic anomaly equation (\ref{BCOV-04-27}), we find a simple
recursion relation for all the cases $N_f=0,1,2,3$ of
Seiberg-Witten theory
\begin{eqnarray}
A^{(2)}&=& \frac{5}{36},\nonumber \\
A^{(g)} &=& (g-1)A^{(g-1)}+\sum_{r=2}^{g-2}A^{(r)}A^{(g-r)},~~g>2
\end{eqnarray}
The first few coefficients $A^{(g)}$ are $\frac{5}{36},\frac{5}{18}, 
\frac{1105}{1296}, \frac{565}{162},\cdots$. 

In the next three subsections we discuss the massless $N_f=1,2,3$ cases 
one by one.

\subsection{$N_f=1$}
For the $N_f=1$ theory the discriminant is according  to (\ref{PFmassless})
\begin{eqnarray}
\Delta_1=16(16u^3+27)\ . 
\end{eqnarray}
The solution of the Picard-Fuchs equation in the  weak coupling limit
$u\rightarrow \infty$ is
\begin{eqnarray}
a&=&
\sqrt{u}(1+\frac{3}{64u^3}-\frac{315}{16384u^6}+\frac{15015}{1048576u^9}+\cdots)
\nonumber \\
a_D &=&
3\, a\,\textrm{log}(u)+\sqrt{u}(\frac{3}{32u^3}-\frac{297}{16384u^6}+\frac{9047}{1048576u^9}+\cdots)
\end{eqnarray}
The prepotential and gauge coupling $\tau$ are then determined by
$\frac{\partial F^{(0)}}{\partial a}=a_D$ and $\tau=-\frac{1}{2\pi
i}\frac{\partial^2 F^{(0)}}{\partial a^2}$. The modulus $u$ can be
expressed in terms of $\tau$ as \cite{Nahm}
\begin{eqnarray} \label{masslessNf1u}
u &=& \frac{1}
{8}(-E_4(\tau)^3-E_6(\tau)E_4(\tau)^{\frac{3}{2}})^{\frac{1}{3}}\eta(\tau)^{-8}
\nonumber \\
&=&\frac{3}{2}m^2\frac{E_4^{\frac{1}{2}}}{(E_4^{\frac{3}{2}}-E_6)
^{\frac{1}{3}}}
\end{eqnarray}
where $m$ is a one-sixth root of $-1$, i.e. $m^6=-1$. 

The holomorphic limit of the genus one amplitude is~\cite{KMT,Huang}
\begin{equation}
F^{(1)}=-\frac{1}{2}\textrm{log}(\frac{d a}{d
u})-\frac{1}{12}\textrm{log}(\Delta_1)\ . 
\label{hf1}
\end{equation}
This follows from $F^{(1)}=-\frac{1}{2} \log(G_{u\bar u} |\Delta|^\frac{1}{6})$, which satisfies 
(\ref{anomaly}), in the holomorphic limit.

To provide modular formulas for all expression we rewrite the Picard-Fuchs equation for $N_f=1$ in
(\ref{PFmassless})  as
\begin{equation} \label{PFNf1perioda}
(4u^2+\frac{27}{4u})[(\frac{d^2a}{d\tau^2})/(\frac{du}{d\tau})^2-(\frac{da}{d\tau})(\frac{d^2u}{d\tau^2})/(\frac{du}{d\tau})^3]+a=0
\end{equation}
Using (\ref{masslessNf1u}) we can obtained a differential equation for period $a$ in 
terms of $\tau$. In the weak coupling limit $\tau\rightarrow i\infty$, the modulus $u$
goes like $u\sim (E_4^{\frac{3}{2}}-E_6)^{-\frac{1}{3}}$, and the
period $a$ goes like $a\sim \sqrt{u}\sim (E_4^{\frac{3}{2}}-E_6)^{-\frac{1}{6}}$.
After fixing the normalization, it follows that the solution of (\ref{PFNf1perioda}) 
that corresponds to the period $a$ is
\begin{equation} \label{Nf1masslessperioda}
a=\sqrt{\frac{3}{2}}\frac{m}{2}\frac{E_4^{\frac{1}{2}}+E_2}{(E_4^{\frac{3}{2}}-E_6)
^{\frac{1}{6}}}\ . 
\end{equation}
As expected the period $a$ has formally modular weight one, since
$\tau=-\frac{1}{2\pi i}\frac{\partial^2 F^{(0)}}{\partial a^2}$
and $F^{(g)}$, $\frac{d}{d\tau}$ have modular weight zero, two
respectively. 

The holomorphic genus one amplitude $F^{(1)}$, i.e. $-\log(\Phi(\tau))$ 
is then
\begin{eqnarray}
F^{(1)}&=& -\frac{1}{12}\textrm{log}(E_4^{\frac{3}{2}}+E_6)\ .
\end{eqnarray}

We integrate the holomorphic anomaly equation (\ref{BCOV-04-27})
and  expand $F^{(g)}$ around the discriminant points
$\Delta_1(u)=0$, in order to use the gap structure. The 3 discriminant points
$\Delta_1(u)=0$ are related by a $Z_3$ symmetry so we only need to
consider the dual expansion around one of the 3 points. 
According to~\cite{SW2} theses points should be related to the weak 
coupling limit $\tau\rightarrow +i\infty$ by an S-duality 
transformation $\tau\rightarrow -\frac{1}{\tau}$. 
The Eisenstein series $E_n$ transform with modular weight $n$, and
a shift for $E_2$, i.e. $E_2\rightarrow \tau^2(E_2+\frac{12}{2\pi i\tau}),~~E_4\rightarrow
\tau^4E_4,~~ E_6\rightarrow \tau^6E_6$. In the weak coupling limit 
$\tau^2$ is negative, so we find that under a S-duality transformation, 
$E_2$, $E_6$ change sign and $E_4$ doesn't. Following the approach 
in~\cite{Huang}, we can find the dual period $a_D$ and $F^{(g)}_D$ by replacing $E_2$,
$E_6$ with $-E_2$, $-E_6$ in (\ref{Nf1masslessperioda})
\begin{equation}
a_D=\sqrt{\frac{3}{2}}\frac{m}{2}\frac{E_4^{\frac{1}{2}}-E_2}{(E_4^{\frac{3}{2}}+E_6)
^{\frac{1}{6}}} \ . 
\end{equation}
Note that the modulus $u$ transforms as
\begin{eqnarray}
u=\frac{3}{2}m^2\frac{E_4^{\frac{1}{2}}}{(E_4^{\frac{3}{2}}-E_6)^{\frac{1}{3}}}
~~\rightarrow~~
\frac{3}{2}m^2\frac{E_4^{\frac{1}{2}}}{(E_4^{\frac{3}{2}}+E_6)^{\frac{1}{3}}},
\end{eqnarray}
i.e. the S-duality transforms $u$ indeed from $u=\infty$ to the
$Z_3$ symmetric discriminant points $u=\frac{3}{2^{\frac{4}{3}}}m^2$.

It is now straightforward to expand the dual genus two amplitude
$F^{(2)}_D$ in terms of the dual period $a_D$ in the weak coupling
limit of the S-dual theory $\tau_D=-\frac{1}{\tau}\rightarrow
+i\infty$. We use the gap condition as in the  case of pure gauge
theory \cite{Huang} to fix the holomorphic ambiguity, and we find
the genus two amplitude and its S-dual
\begin{eqnarray} \label{Nf1masslessF2}
F^{(2)}&=&
\frac{(E_4^{\frac{3}{2}}-E_6)^{\frac{1}{3}}}{2160m^2(E_4^{\frac{3}{2}}+E_6)^2}
[ -25E_2^3E_4+E_2^2(-135E_4^{\frac{3}{2}}+30E_6)\nonumber \\ &&
+E_2(255E_4^2-120E_4^{\frac{1}{2}}E_6)
-159E_4^{\frac{5}{2}}+140E_4E_6]\\
 F^{(2)}_D&=&
\frac{(E_4^{\frac{3}{2}}+E_6)^{\frac{1}{3}}}{2160m^2(E_4^{\frac{3}{2}}-E_6)^2}
[ 25E_2^3E_4+E_2^2(-135E_4^{\frac{3}{2}}-30E_6)\nonumber \\
&& -E_2(255E_4^2+120E_4^{\frac{1}{2}}E_6)
 -159E_4^{\frac{5}{2}}-140E_4E_6]
\end{eqnarray}
The genus two space-time instanton expansion and the S-dual
expansion are as follows
\begin{eqnarray}
F^{(2)}&=&\frac{1}{160a^2}+\frac{9}{1024a^8}-\frac{16749}{262144a^{14}}
+\frac{187215}{1048576a^{20}}\ ,\nonumber
\\ &&
-\frac{6536606985}{17179869184a^{26}}+\mathcal{O}(\frac{1}{a^{32}})
\\
F^{(2)}_D&=&-\frac{1}{240a_D^2}-\frac{221a_D}{3^{\frac{1}{2}}62208m^3}
+\frac{76289a_D^2}{2^{\frac{1}{3}}10077696m^4}
-\frac{1082609a_D^3}{2^{\frac{2}{3}}3^{\frac{1}{2}}45349632m^5}+\mathcal{O}(a_D^4)\ . 
\nonumber \\
\end{eqnarray}

We obtain the genus three amplitude using the gap condition at the conifold point
\begin{eqnarray}
F^{(3)}&=&\frac{(E_4^{\frac{3}{2}}-E_6)^{\frac{2}{3}}}{544320m^4(E_4^{\frac{3}{2}}+E_6)^4}
\{525E_2^6E_4^2-350E_2^5(19E_4^{\frac{5}{2}}-5E_4E_6)\nonumber \\
&&
+35E_2^4(1225E_4^3-694E_4^{\frac{3}{2}}E_6+16E_6^2)-280E_2^3(637E_4^{\frac{7}{2}}-546E_4^2E_6
+51E_4^{\frac{1}{2}}E_6^2) \nonumber \\&&
+7E_2^2(67221E_4^4-75400E_4^{\frac{5}{2}}E_6+14540E_4E_6^2)
\nonumber \\ &&
-14E_2(49821E_4^{\frac{9}{2}}-68867E_4^3E_6+20960E_4^{\frac{3}{2}}E_6^2-560E_6^3)
\nonumber \\ &&
+(440325E_4^5-720006E_4^{\frac{7}{2}}E_6+308700E_4^2E_6^2-22400E_4^{\frac{1}{2}}E_6^3)\ ,
 \}
\end{eqnarray}
which yields to lowest order in the asymptotically free region and near the conifold
\begin{eqnarray} \label{F3masslessNf1instanton}
F^{(3)}&=&
\frac{5}{2688a^4}-\frac{3}{1024a^{10}}+\frac{96453}{524288a^{16}}
-\frac{6065417}{4194304a^{22}} \nonumber \\ &&
+\frac{213776429067}{34359738368a^{28}}+\mathcal{O}(\frac{1}{a^{34}})
\\ F^{(3)}_D &=&
\frac{1}{1008a_D^4}-\frac{197a_D}{2^{\frac{2}{3}}3^{\frac{1}{2}}165888m^5}
-\frac{54542723a_D^2}{19591041024}\nonumber \\ &&
+\frac{159862731109
a_D^3}{2^{\frac{1}{3}}3^{\frac{1}{2}}9873884676096m}+\mathcal{O}(a_D^4)\ .
\end{eqnarray}
The instanton expansion (\ref{F3masslessNf1instanton})
agrees with Nekrasov's calculations (\ref{NekrasovmasslessNf1})
and makes predictions at higher instanton numbers.

\subsection{$N_f=2$}
The discriminant is
\begin{eqnarray}
\Delta_2=(4u^2-1)^2\ .
\end{eqnarray}
That it is of fourth order in $u$ can be seen  
from (\ref{conifolddivisors}) in a later section, where we provide the 
expression of the conifold divisor for generic flavor masses. 
The solution of the Picard-Fuchs equation at weak coupling limit
$u\rightarrow \infty$ is
\begin{eqnarray}
a&=&
\sqrt{u}(1-\frac{1}{64u^2}-\frac{15}{16384u^4}-\frac{105}{1048576u^6}+\cdots)
\nonumber \\
a_D &=&
2a\textrm{log}(u)+\sqrt{u}(-\frac{1}{32u^2}-\frac{13}{16384u^4}-\frac{163}{3145728u^6}+\cdots)
\end{eqnarray}
The modulus $u$ can be expressed in terms of $\tau=-\frac{1}{2\pi
i}\frac{\partial^2 F^{(0)}}{\partial a^2}$ as \cite{Nahm}
\begin{eqnarray} \label{Nf2u}
u=\frac{1}
{16}\frac{\eta(\frac{\tau}{2})^8}{\eta(2\tau)^8}+\frac{1}{2}=\frac{\theta_4^4(\tau)}{\theta_2^4(\tau)}+\frac{1}{2}\ .
\end{eqnarray}
We verify the genus one amplitude satisfy the holomorphic anomaly
equation
\begin{eqnarray} \label{Nf2F1}
F^{(1)}=-\frac{1}{2}\textrm{log}(\frac{d a}{d
u})-\frac{1}{12}\textrm{log}(\Delta_2)\ .
\label{F1nf2} 
\end{eqnarray}

The period $a$ can also be written in terms of theta functions of
$\tau$. We notice that the Picard-Fuchs equation for $N_f=2$ is very
similar to the one of $N_f=0$ pure Seiberg-Witten theory studied in
\cite{Huang}. In particular, if we change the normalization
$u\rightarrow \frac{u}{2}$, but leave $\tau$ and $a$ fixed, then
the Picard-Fuchs equation (\ref{PFmassless}) and the expression of
$u$ (\ref{Nf2u}) are exactly the same as that of pure $SU(2)$
theory. So the expression of the period $a$ in terms of $\tau$ should
be the same as that of \cite{Huang} up to multiplicative constant.
We find
\begin{eqnarray} \label{a-05-19}
a=\frac{1}{3\theta_2^2(\tau)}(E_2(\tau)+\theta_3^4(\tau)+\theta_4^4(\tau))\ .
\end{eqnarray}
However since $\Delta_2$ is the square of the $N_f=0$ discriminate $\Delta_1$. 
Therefore the genus one amplitude is not simply $F^{(1)}=-\textrm{log}(\eta(\tau))$ 
as in the $N_f=0$ case~\cite{Huang} but rather
\begin{eqnarray} \label{F1-05-19}
F^{(1)}=-\frac{1}{3}\textrm{log}(\frac{\theta_3^2(\tau)\theta_4^2(\tau)}{\theta_2(\tau)})\ .
\end{eqnarray}
In the following we use the notation in~\cite{Huang} and define
\begin{equation}
\label{defabcd}
b:=\theta_2^4(\tau), \qquad c:=\theta_3^4(\tau)=b+d,\qquad d:=\theta_4^4(\tau)\ ,
\end{equation} 
$h:=b+2d$ and  $X:=\frac{b}{1728c^2d^2}$.

Under a $S$ duality transformation, the theta functions have modular
weight two, and transform as $b\rightarrow -\tau^2d$,
$c\rightarrow -\tau^2c$, $d\rightarrow -\tau^2b$. The Eisenstein
$E_2$ is weight two and transforms with a shift (\ref{E2trans}). 
The period $a$ contains $E_2$ as well as $\theta_2^2$ and is therefore not 
modular invariant under $\Gamma(2)$. By a duality transformation followed 
by a holomorphic limit it is rather related to the dual period
\begin{eqnarray}
a_D=-\frac{i}{3\theta_4^2}(E_2-b-c)\ .
\end{eqnarray}

Using the gap condition from dual expansion, we fix the genus two
amplitude and find the space-time instanton expansion
\begin{eqnarray} \label{Nf2masslessF2}
F^{(2)}=
\frac{2}{15}X\{25E_2^3-75E_2^2h+15E_2(13b^2+22cd)-h(137b^2+8cd)\}\ ,
\end{eqnarray}
\begin{eqnarray}
F^{(2)}&=&\frac{7}{480a^2}-\frac{7}{1024a^6}-\frac{1425}{262144a^{10}}
 -\frac{15717}{8388608a^{14}} \nonumber
\\ &&
-\frac{8623029}{17179869184a^{18}}+\mathcal{O}(\frac{1}{a^{22}})\ .
\end{eqnarray}
The S-dual of $F^{(2)}$ and its dual expansion fulfilling the gap 
condition are
\begin{eqnarray}
F^{(2)}_D=
\frac{2}{15}X_D\{25E_2^3-75E_2^2h_D+15E_2(13d^2+22bc)-h_D(137d^2+8bc)\}\ .
\end{eqnarray}
\begin{eqnarray}
F^{(2)}_D=\frac{1}{120a_D^2}+\frac{3ia_D}{512}-\frac{33a_D^2}{1024}
-\frac{2147ia_D^3}{20480}+\mathcal{O}(a_D^4)\ ,
\end{eqnarray}
where $h_D=-d-2b$, $X_D=-\frac{d}{1728b^2c^2}$.

We push the analysis to genus three using the gap condition. The
genus three amplitude is
\begin{eqnarray}
F^{(3)}&=& X^2\{80E_2^6-480E_2^5h+48E_2^4(41b^2+104cd)
-\frac{32}{3}E_2^3h(646b^2+685cd) \nonumber \\ &&
+\frac{16}{5}E_2^2(6503b^4+23410b^2cd+7637c^2d^2)  \nonumber \\ &&
-\frac{32}{5}E_2h(5867b^4+11605b^2cd+671c^2d^2) \nonumber \\ &&
+\frac{16}{105}(177293b^6+787182b^4cd+619233b^2c^2d^2+40232c^3d^3)\}\ .
\end{eqnarray}
The space-time instanton expansion and the S-dual expansion are given below
\begin{eqnarray} \label{F3masslessNf2instanton}
F^{(3)}&=&
\frac{31}{8064a^4}+\frac{5}{2048a^8}+\frac{8843}{524288a^{12}}
+\frac{140721}{8388608a^{16}} \nonumber \\ &&
+\frac{318316439}{34359738368a^{22}}+\mathcal{O}(\frac{1}{a^{26}})
\\ F^{(3)}_D &=&
\frac{1}{504a_D^4}+\frac{45ia_D}{16384} -\frac{279a_D^2}{8192}
-\frac{745933ia_D^3}{3670016}+\mathcal{O}(a_D^4)\ .
\end{eqnarray}
The instanton expansion (\ref{F3masslessNf2instanton}) agree
with (\ref{NekrasovmasslessNf2}) and make predictions at 
higher instanton numbers.

\subsection{$N_f=3$}

The discriminant is
\begin{eqnarray}
\Delta_3=\frac{u}{4}(16u-1)\ . 
\end{eqnarray}
The solution of the Picard-Fuchs equation at weak coupling limit
$u\rightarrow \infty$ is
\begin{eqnarray}
a&=&
\sqrt{u}(1-\frac{1}{64u}-\frac{3}{16384u^2}-\frac{5}{1048576u^3}+\cdots)
\nonumber \\
a_D &=&
a\textrm{log}(u)+\sqrt{u}(-\frac{1}{32u}-\frac{1}{16384u^2}+\frac{1}{3145728u^3}+\cdots)\ . 
\end{eqnarray}
In the case $N_f=3$, it turns out to be convenient to define the
gauge coupling as $\tau=-\frac{1}{\pi i}\frac{\partial^2
F^{(0)}}{\partial a^2}+1$. The modulus $u$ can be expressed in
terms of $\tau$ as \cite{Nahm}
\begin{eqnarray} \label{Nf3u}
u=-\frac{1}
{256}\frac{\eta(\frac{(\tau-1)}{2})^8}{\eta(2(\tau-1))^8}=\frac{1}
{256}\frac{\eta(\frac{\tau}{2})^8}{\eta(2\tau)^8}+\frac{1}{16}
\end{eqnarray}
We verify that the genus one amplitude satisfy the holomorphic anomaly
equation
\begin{eqnarray}
F^{(1)}=-\frac{1}{2}\textrm{log}(\frac{d a}{d
u})-\frac{1}{12}\textrm{log}(\Delta_3)-\frac{1}{4}\textrm{log}(u)\ .
\end{eqnarray}
We note an additional singularity at $u=0$ besides the
$\Delta_3=0$.

We can also use the results from pure $SU(2)$ case \cite{Huang} to
write period $a$ and $F^{(1)}$ as theta functions of $\tau$. We
notice that under a change of variable $u=\frac{\tilde{u}+1}{32}$,
the Picard-Fuchs equation (\ref{PFmassless}) and the formula
(\ref{Nf3u}) become the same as the pure gauge theory case, namely
we have
\begin{eqnarray}
4(\tilde{u}^2-1)\frac{d^2a}{d\tilde{u}^2}+a &=& 0 \nonumber \\
\tilde{u}=\frac{1}{8}\frac{\eta(\frac{\tau}{2})^8}{\eta(2\tau)^8}+1
&=&\frac{\theta_3^4(\tau)+\theta_4^4(\tau)}{\theta_2^4(\tau)}\ .
\nonumber
\end{eqnarray}
So we can use the result of pure $SU(2)$ gauge theory and we find
\begin{eqnarray}
a=\frac{1}{12\theta_2^2(\tau)}(E_2(\tau)+\theta_3^4(\tau)+\theta_4^4(\tau))
\end{eqnarray}
and the genus one amplitude
\begin{eqnarray}
F^{(1)}&=&-\frac{1}{3}\textrm{log}(\frac{\theta_3^4(\tau)\theta_4(\tau)}{\theta_2^2(\tau)})\ . 
\end{eqnarray}

In the case of $N_f=3$, there are two different dual expansions:
one at $u=0$ and one at $u=\frac{1}{16}$. Unlike the case in
$N_f=0$ and $N_f=2$, the two expansions are not related by a $Z_2$
symmetry. The S-duality transformation transforms $u=\infty$ to
$u=\frac{1}{16}$ or $\tilde{u}=1$, while a T and S duality
transforms $u=\infty$ to $u=0$ or $\tilde{u}=-1$. The corresponding actions 
on $a,b$ and $c$ are
\begin{eqnarray}
\textrm{S-duality}:&&~~~ b\rightarrow -\tau^2d, c\rightarrow
-\tau^2c, d\rightarrow -\tau^2b
\nonumber \\
\textrm{TS-duality}: &&~~~ b\rightarrow \tau^2d, c\rightarrow
-\tau^2b, d\rightarrow -\tau^2c \ .\nonumber
\end{eqnarray}
and yield the dual periods as
\begin{eqnarray}
\textrm{S-duality}: &&~~~a_{D1}=-\frac{i}{12\theta_4^2}(E_2-b-c)
\nonumber
\\
\textrm{TS-duality}: &&~~~a_{D2}=-\frac{1}{12\theta_4^2}(E_2-b-c)\ .
\nonumber
\end{eqnarray}
It turns out there are gap structures in the dual series
expansions at both $u=0$ and $u=\frac{1}{16}$, where the first
sub-leading terms of the dual series at $u=0$ and $u=\frac{1}{16}$
go like constant and $a_D^3$ respectively. We are able to use this
structure to fix the genus two amplitude
\begin{eqnarray}
F^{(2)}&=&\frac{b}{810c^2d^2}[50E_2^3-90E_2^2(b+4d)+30E_2(2b^2-4bd+35d^2)
\nonumber \\ && -(16b^3+51b^2d-1428bd^2+443d^3)]\ .
\end{eqnarray}
The space-time instanton expansion and the dual expansions are
\begin{eqnarray}
F^{(2)}&=&
\frac{11}{480a^2}+\frac{5}{1024a^4}-\frac{109}{262144a^6}+\frac{83}{4194304a^8}
\nonumber \\ &&
-\frac{13361}{1717869184a^{10}}+\mathcal{O}(\frac{1}{a^{12}})\ ,
\\
F^{(2)}_{D1} &=&
\frac{1}{240a_{D1}^2}+\frac{1}{2}+\frac{271ia_{D1}}{16}-\frac{3811a_{D1}^2}{16}
-\frac{50781ia_{D1}^3}{20}+\mathcal{O}(a_{D1}^4)\ ,
\\
F^{(2)}_{D2} &=&
\frac{1}{15a_{D2}^2}+\frac{4a_{D2}^3}{5}-\frac{75a_{D2}^4}{4}+\frac{2155a_{D2}^5}{8}+\mathcal{O}(a_{D2}^6)\ .
\end{eqnarray}
The vanishing of the subleading coefficients in both of the dual series expansions 
up order $a_{D1}^0$ and $a_{D2}^3$ respectively yield independent conditions. 
These enable us to precisely to fix the unknowns in the ansatz for the 
holomorphic ambiguity, which is increased w.r.t. to the $N_f=0$ and $N_f=2$ case, 
because there is no $Z_2$ symmetry $u\rightarrow -u$ in the $N_f=3$ theory.

We  push the analysis to genus three
\begin{equation}
\begin{array}{rl}
F^{(3)}=&\ds{\frac{2b^2}{76545c^4d^4}}\bigr\{1050E_2^6-1050E_2^5(5b+13d)+210E_2^4(80b^2+233bd+374d^2)-\\[3 mm]
& 70E_2^3(529b^3+1752b^2d+1731bd^2+3764d^3)+42E_2^2(1172b^4+3985b^3d+\\ 
& 4563b^2d^2-9299bd^3+12818d^4) -42E_2(844b^5+3109b^4d+4765b^3d^2+\\ 
& 12404b^2d^3-63022bd^4+9554d^5)+(10718b^6+44304b^5d+81507b^4d^2+\\ 
& 19406b^3d^3+2674506b^2d^4-2382348bd^5+117557d^6)\bigr\}\ .
\end{array}
\end{equation}
The space-time instanton expansion and the two dual expansions are
\begin{eqnarray} \label{Nf3F3series}
F^{(3)}&=&
\frac{47}{8064a^4}-\frac{1}{512a^6}+\frac{769}{534288a^8}-\frac{1595}{8388608a^{10}}
\nonumber \\ &&
+\frac{506627}{34359738368a^{12}}+\mathcal{O}(\frac{1}{a^{14}})\ ,
\\
F_{D1}^{(3)} &=&
\frac{1}{1008a_{D1}^4}+10+\frac{29327ia_{D1}}{32}-\frac{413345a_{D1}^2}{16}
+\mathcal{O}(a_{D1}^3)\ , \\
F_{D2}^{(3)} &=&
\frac{4}{63a_{D2}^4}-\frac{90a_{D2}^3}{7}+\frac{4183a_{D2}^4}{8}-\frac{35483a_{D2}^5}{3}+\mathcal{O}(a_{D2}^6)\ ,
\end{eqnarray}
where the space-time instanton expansion (\ref{Nf3F3series}) again
agrees with Nekrasov's calculations (\ref{NekrasovmasslessNf3})
and makes predictions at higher instanton numbers.  Note that the
non-zero sub-leading term of the two dual series appear at order  
$a_{D1}^0$ and $a_{D2}^3$ respectively, as the genus two case.

\section{$SU(2)$ Seiberg-Witten theory with massive
hypermultiplets} \label{sectionmassive}

In this section, we will show that the gravitational couplings
$F^{(g)}$ for the massive Seiberg-Witten theory can be solved 
as a polynomials of generators of the relevant modular functions, 
whose coefficients are rational functions of the Coulomb modulus $u$ 
as well as the mass parameters $m_i$. 
The equation $J(\tau)=R_{N_F}(u,\underline{ m})$ still governs the occurrence 
of the modular functions and relevant group is again the quotient of 
${\rm PSL}(2,\mathbb{Z})$ by the permutation group acting on the 
roots $u(J,\underline{m})$. Mathematically the mass parameters $m_i$ are 
known as isomonodromic deformation parameters.

\subsection{The prepotential $F^{(0)}$}
For the massive case the Picard-Fuchs equation is much more
complicated than the massless case. There is a standard though
tedious way to derive the Picard-Fuchs equations from the
Seiberg-Witten curve \cite{KLT}. The Picard-Fuchs equation for
$SU(2)$ Seiberg-Witten theory with massive hypermultiplets were
derived in \cite{Ohta1, Ohta2} and for the case of one
massive hypermultiplet ($N_f$=1) it is given by 
\begin{eqnarray} \label{PF1}
&& \frac{d^3\Pi}{du^3}+\frac{3\Delta_1+(4m_1^2-3u)\frac{\partial
\Delta_1}{\partial m_1}}{\Delta_1(4m_1^2-3u)}\frac{d^2\Pi}{du^2}
\nonumber
\\
&&
-\frac{8[4(2m_1^2-3u)(4m_1^2-3u)+3(3\Lambda_1^3m_1-4u^2)]}{\Delta_1(4m_1^2-3u)}\frac{d\Pi}{du}=0 \ .
\end{eqnarray}
Here $m_1$ is the mass of the hypermultiplet and $\Delta_1$ is
the discriminant of the Seiberg-Witten curve
\begin{equation}
\Delta_1=256u^3-256m_1^2u^2-288\Lambda_1^3m_1u+256\Lambda_1^3m_1^3+27\Lambda_1^6\ .
\end{equation}
The differential equation~\ref{PF1} has a second singularity at the vanishing locus of
\begin{equation}
\Delta_2=4 m_1^2-3u \ .
\end{equation} 
In order to match with Nekrasov's convention \cite{Nekrasov1}, we
again set the dynamical scale $\Lambda_1=2^{\frac{2}{3}}$.

In the weak coupling region near $u\rightarrow\infty$, the Picard-Fuchs
equation (\ref{PF1}) has a constant solutions and two other
solution corresponding to the periods $a$ and $a_D$. It was found
in \cite{Ohta1}
\begin{eqnarray}
a&=& \sqrt{u}(1-\frac{m_1}{4u^2}+\frac{3}{64u^3}-
\frac{15m_1^2}{64u^4}+\frac{35m_1}{256u^5}-\frac{105(3+64m_1^3)}{16384u^6} \nonumber \\
&& +\frac{3465m_1^2}{8192u^7}+\mathcal{O}(\frac{1}{u^{8}})\ ,
\nonumber \\
a_D&=&
3a\textrm{log}(u)+\sqrt{u}[-\frac{m_1^2}{u}-(\frac{m_1}{2}+\frac{m_1^4}{6})\frac{1}{u^2}
+(\frac{3}{32}-\frac{m_1^3}{4}-\frac{m_1^6}{15})\frac{1}{u^3}
\nonumber \\
&&
+(\frac{3m_1^2}{64}-\frac{m_1^5}{8}-\frac{m_1^8}{28})\frac{1}{u^4}+\mathcal{O}(\frac{1}{u^5})]\ .
\end{eqnarray}
We solved the prepotential using $\frac{\partial
F^{(0)}}{\partial a}=a_D$ and checked it against  Nekrasov's result summarized in
(\ref{NekrasovF0}).

There are relations between the period $a$, Coulomb
modulus $u$ and the gauge coupling $\tau$, which become useful for solving the model.  
After a $PGL(3,\mathbb{C})$ transformation, the massive  Seiberg-Witten curve  
can brought into  Weierstrass form (\ref{weierstrassform}). For $N_f=1$ one has 
\begin{equation} \label{gmassive} 
\begin{array}{rl}
g_2(u,m_1)=& \frac{4}{3}u^2-4m_1 \\
g_3(u,m_1)=& -\frac{8}{27}u^3+\frac{4}{3}m_1u-1\ .
\end{array}
\end{equation}  

Using the fact that the period equation is solved in terms of modular forms~\cite{Fricke} (see 
\cite{zagier} for an elementary review), the definition of $\tau=-\frac{1}{2\pi i}\frac{\partial a_D}{\partial a}$ 
and the $J$ function (\ref{J}) one can show that the period $a$ satisfies the 
equation~\cite{Brandhuber} 
\begin{equation} \label{aumassiveNf1}
\frac{d u}{d a}=\sqrt{-18\frac{g_3(u,m_1)}{g_2(u,m_1)}\frac{E_4(\tau)}{E_6(\tau)}}\ .
\end{equation}
Note that this equation is universal for $N_f=0,1,2,3$ in the sense that it depends on the 
specifics of the curve only via $g_2(u,\underline{m})$  and $g_3(u,\underline{m})$.

\subsection{Gravitational couplings}

We shall follow the approach in \cite{Huang} and treat the period
$a$ as a flat coordinate in the holomorphic limit. The singular locus of 
the Picard-Fuchs equation (\ref{PF1}) is at $\Delta_1=0$ and $\Delta_2=0$. 
$\Delta_1=0$ is the the conifold divisor, i.e. one hypermultiplet becomes 
massless for these values of the moduli. According to~\cite{Vafa:1995ta} each 
hypermultiplet contributes $-\frac{1}{12}$ to one-loop gravitational $\beta$ 
function, which yields a logarithmic running of the coupling of $R_+^2$.  
This gives rise to an $-\frac{1}{12} \log(\Delta_1) R_+^2$ contribution 
in the one loop effective action, compare (\ref{gravitationalcorrections}). 
On the other hand the conformal locus lies on $\Delta_2=0$, as discussed further in 
section \ref{sectionsuperconfomal}, and here the $\beta$ function and hence the $\log(\Delta_2)$ contribution to $F^{(1)}$ 
vanishes. This fixes the ambiguity at genus zero and the holomorphic limit~\cite{KMT} 
\begin{eqnarray} \label{Nf1genus1}
F^{(1)}=-\frac{1}{2}\textrm{log}(\frac{d a}{d
u})-\frac{1}{12}\textrm{log}(\Delta_1)
\end{eqnarray}
agrees with Nekrasov's calculation (\ref{NekrasovF1}) up to
an ambiguous constant. The form of (\ref{Nf1genus1}) has been already 
noted in~\cite{Eguchi}. 

Using (\ref{aumassiveNf1}) the genus one gravitational correction $F^{(1)}$
can be also written as
\begin{eqnarray} \label{C109}
F^{(1)}=-\frac{1}{12}\textrm{log}(\frac{g_2^3(g_2^3-27g_3^2)}{g_3^3}\frac{E_6^3}{E_4^3})\ .
\end{eqnarray}

As discussed in section (\ref{sectionmasslessdirect}) the $F^{(g)}$ ($g\geq 2$) of Seiberg-Witten theory are modular
invariant with weight zero. As we have seen the covariant 
derivatives in the recursion lead to an an-holomorphic dependence, 
which can be completely absorbed into powers of the non-holomorphic 
Eisenstein series $\hat{E}_2(\tau)$. Their modular 
transformation must  be compensated by holomorphic modular forms. We claim that 
that all an-holomorphic dependence can be absorbed into the 
weight zero an-holomorphic modular form
\begin{eqnarray} \label{definitionX}
X(\tau)=\frac{\hat{E}_2(\tau)E_4(\tau)}{E_6(\tau)}\ .
\end{eqnarray}
This can be established by rewriting the holomorphic anomaly equation ($g\geq 2$) as
\begin{eqnarray} \label{C111}
\frac{\partial F^{(g)}(X,u)}{\partial X} =
\frac{E_6}{24E_4}(\frac{d^2
F^{(g-1)}}{da^2}+\sum_{r=1}^{g-1}\frac{d F^{(r)} }{da}\frac{d
F^{(g-r)}}{da})\ .
\end{eqnarray}
Using the relations of $a$, $u$ and $\tau$ in (\ref{J}) (\ref{aumassiveNf1}), 
and induction one can easily show the right hand side of the above equation 
is a polynomial of $X$ of degree $3g-4$ whose coefficients as rational 
function of $u$, consistent with the induction. It follows that the higher 
genus $F^{(g)}$ ($g\geq 2$) are polynomials of $X(\tau)$ of degree\footnote{As we explained in the
massless cases in Sec. \ref{sectionmasslessdirect}, there is an
isomorphism between $E_2$ and $\hat{E}_2$. So one treat the above
equations as in the holomorphic limit by replacing $\hat{E}_2$ 
with $E_2$.} $3g-3$, whose coefficients are rational functions 
of the modulus $u$ and $m_1$.

Given (\ref{C109}) it is easy to integrate (\ref{C111}) up to 
the holomorphic ambiguity. E.g. for genus two we find
\begin{equation} 
\begin{array}{rl} \label{C112}
F^{(2)} &=
\frac{1}{\Delta(u)^2} \bigl\{-2880(3u-4m_1^2)^2\left(\frac{g_3X}{g_2}\right)^3-96(252u^3-648m_1^2u^2+(352m_1^4+\\[2 mm]
&54m_1)u+27(8m_1^3-9))\left(\frac{g_3X}{g_2}\right)^2 -\frac{64}{3}[324u^4-528m_1^2u^3+4m_1(76m_1^3-27)u^2-\\ 
&-36(26m_1^3+27)u+3m_1^2(128m_1^3+729)]\left(\frac{g_3X}{g_2}\right) \}+ f^{(2)}\ ,
\end{array}
\end{equation}
where $f^{(2)}$ is the holomorphic ambiguity, which is again a rational
function of $u$ and $m_1$.

\subsection{Fixing the holomorphic ambiguity with dual expansions}

In order to fix the holomorphic ambiguity, we use the gap
structure in the dual expansion at a discriminant point $u=u_1$,
where $\Delta(u_1)=0$. The formulae are essentially the same as
the semi-classical limit $u\rightarrow \infty$, and we just need
to use the dual coupling and period $\tau_D$ and $a_D$ in places
of $\tau$, $a$. The formulae (\ref{J}),
(\ref{aumassiveNf1}) become
\begin{eqnarray} \label{3.77-10-31-2008}
J(\tau_D)=\frac{g_2(u)^3}{g_2(u)^3-27g_3(u)^2}  \\
\frac{d u}{d
a_D}=\sqrt{-18\frac{g_3(u)}{g_2(u)}\frac{E_4(\tau_D)}{E_6(\tau_D)}}
\end{eqnarray}
{}From the above formulae we see that around the discriminant point
$u\sim u_1$, the dual theory is indeed weakly coupled in terms of
the dual coupling $\tau_D\rightarrow i\infty$, and the asymptotic
behavior of the dual period is $a_D\sim (u-u_1)$.

We can now replace $\tau$ in the genus two formula (\ref{C112})
with dual coupling $\tau_D$, and expand $F^{(2)D}$ in terms of the
dual period $a_D$. 

The ambiguity $f^{(g)}$ can a priori have poles at the singular 
points of Picard-Fuchs equation $\Delta(u)=0$ and $\Delta_2=0$. 
However $\Delta_2=0$ does not appear as a pole in the holomorphic
ambiguity. This is because there is a conformal massless spectrum 
at that locus in the moduli space, similar to the situation encountered 
in~\cite{HKQ} for the Gepner point in the Calabi-Yau moduli space.
The singular behavior of $f^{(g)}$ at the conifolds 
implies then that  $f^{(g)}=\frac{p_{n}(u)}{\Delta_1^{2g-2}}$, where $p_n(u)$ is a
polynomial in $u$ of degree $n$. Since $f^{(g)}$ must be regular at 
$u\rightarrow \infty$ and $\Delta_1\sim u^3$ we get that $n\le 6g-6$. In fact
it turns out that  $n= 6g-7$. Using the $2g-1$ first coefficients in (\ref{gapm}) 
at the three inequivalent conifold points, we see that the equations following
from the gap condition overdetermine the coefficients of $p_n(u)$. This fixes
the holomorphic ambiguity for all $g$. E.g. for genus two we get
\begin{equation} \label{C116}
\begin{array}{rl}  
f^{(2)}&=\frac{128}{405\Delta(u)^2}[-684u^5+3192m_1^2u^4-2m_1(656m_1^3+4293)u^3 +378(8m_1^3+45)u^2 -\\ & 
54m_1^2(80m_1^3+183)u+27(664m_1^3-729)] \ .
\end{array}
\end{equation}
Unlike the massless case, here the constant term in the dual
expansion does not vanish. Although it is too complicated to write
down the constant term, we have checked it indeed vanishes in the
massless limit $m_1=0$.

The holomorphic ambiguity (\ref{C116}) together with (\ref{C112})
gives the exact formula for genus two $F^{(2)}$ in the massive
$N_f=1$ Seiberg-Witten theory. We have checked the agreement with
Nekrasov's instanton calculation in semi-classical limit. We can
furthermore obtain predictions for higher instanton results at
genus two. For example, the 6-instanton and 7-instanton results
are
\begin{eqnarray}
F^{(2)}_{6-instanton} &=&
\frac{1497720a^6-39720651a^4m_1^2+131881442a^2m_1^4-96877135m_1^6}{8388608a^{26}}
\nonumber \\
F^{(2)}_{7-instanton} &=&
\frac{3(6542298a^6m_1-73190615a^4m_1^3+181612908a^2m_1^5-117503791m_1^7)}
{8388608a^{30}}\ . \nonumber
\end{eqnarray}

\subsection{Comments on the $N_f=2,3$ cases and integrability}

We can transform the Seiberg-Witten curve for $Nf=2,3$ with generic hypermultiplet masses 
into Weierstrass  form. The conifold divisor $\Delta=g_2^3-27g_3^2=0$ for the $N_f=2,3$ 
cases are (here we use the convention for the QCD scale $\Lambda_{N_f=2}=2$, $\Lambda_{N_f=3}=4$)
\begin{eqnarray}
\Delta_{N_f=2} &=& 16u^4-16(m_1^2+m_2^2)u^3+(16m_1^2m_2^2-80m_1m_2-8)u^2
\nonumber \\ &&  +[36(m_1^2+m_2^2)+72m_1m_2(m_1^2+m_2^2)] u
 \nonumber \\ && +1-12m_1m_2-6m_1^2m_2^2-64m_1^3m_2^3-27(m_1^4+m_2^4)\ ,
 \nonumber \\ \Delta_{N_f=3} &=& -16u^5 +(1+16m_1^2+16m_2^2+16m_3^2)u^4 
 \nonumber \\ && +[8(m_1^2+m_2^2+m_3^2)-16(m_1^2m_2^2+m_1^2m_3^2+m_2^2m_3^2)+88m_1m_2m_3)u^3
 \nonumber \\  &&+ f_2(m_1,m_2,m_3)u^2 + f_1(m_1,m_2,m_3)u++ f_0(m_1,m_2,m_3)\ , 
\label{conifolddivisors}
 \end{eqnarray}  
where $f_2, f_1, f_0$ are some symmetric polynomial of $m_1, m_2,m_3$.  

The number of conifold point is $n=3,4,5$ for $N_f=1,2,3$ and these points are distinct 
for generic hypermultiplet masses. The ambiguity at genus $g$ is a 
rational function whose denominator is $\Delta^{2g-2}$, while regularity at $u=\infty$ constrains 
the number of unknown constants in the holomorphic ambiguity to be 
$n(2g-2)$. The gap boundary conditions at each of the $n$ distinct conifold 
singularities provide $2g-2$ conditions. This is exactly enough information 
to fix the holomorphic ambiguity.

We have also checked the genus one formula 
\begin{eqnarray} 
F^{(1)}=-\frac{1}{2}\textrm{log}(\frac{d a}{d
u})-\frac{1}{12}\textrm{log}(\Delta)
\end{eqnarray}
agrees with Nekrasov's instanton counting formulae for $N_f=2,3$ cases for generic masses when we expand it at $u=\infty$.

\section{$SU(2)$ Seiberg-Witten theory at superconformal points}

\label{sectionsuperconfomal}

One of the most interesting aspect of the Seiberg-Witten solution 
of $N=2$ gauge theories is that it allows to study  regions in the 
parameter spaces where previously unknown types of four dimensional 
interacting field theories have been discovered. Of particular interest 
are the points where several dyons become massless, which have 
electric as well as magnetic charges. It is well known that no 
effective action with only local interactions can be written 
down in that case and one says the massless dyons are mutually 
non-local. Geometrically the situation is characterized by the fact
that pairs of cycles which mutually non-vanishing intersection numbers 
vanish. It also implies that the divisors, where mutually non-local dyons 
vanish, intersect in the moduli space.     

In the previous Section~\ref{sectionmassless} we studied the case
where the bare masses of hypermultiplets are zero. It this case the 
extra massless particles at the colliding singularities are
mutually local.  In \cite{Argyres}, some special cases of
hypermultiplet masses are studied where two mutually non-local
singular points in the $u$ plane collide. In the case there is a
non-trivial interacting superconformal field theory at the
colliding singularity in the $u$ plane. Besides the known $N_f=4$
superconformal field theory, three new $\mathcal{N}=2$
superconformal field theories were found \cite{Argyres} from
$SU(2)$ Seiberg-Witten theory with $N_f=1,2,3$ flavors, and are
denoted as $(N_f,1)$ superconformal field theory respectively .

A technically interesting aspect is that the equation (\ref{J}), 
which is for general masses, compare (\ref{gmassive}), not easily 
solvable for $u(\tau)$, becomes simple and solvable at the 
conformal points, which allows below to find explicit formulas 
for the amplitudes in terms of modular forms.

\subsection{$N_f=1$, $m_1=\frac{3\Lambda_1}{4}$}

We follow our previous notation $\Lambda_1=2^{\frac{2}{3}}$. In the special case of the mass of the flavor
$m_1=\frac{3\Lambda_1}{4}$, there
is a non-trivial $(1,1)$ superconformal field theory at
$u=\frac{3}{2^{\frac{2}{3}}}$ where two mutually non-local
massless dyons collide, and there is another dyon singularity at
$u=-\frac{15}{4\cdot 2^{\frac{2}{3}}}$. The $(1,1)$ superconformal
field theory at
$u=\frac{3}{2^{\frac{2}{3}}}$ is equivalent to the Argyres-Douglas point originally
found in pure $SU(3)$ Seiberg-Witten theory in
\cite{Argyres:1995jj}. The discriminant is
\begin{eqnarray}
\Delta\sim(u-\frac{3}{2^{\frac{2}{3}}})^2(u+\frac{15}{4\cdot
2^{\frac{2}{3}}})\ .
\end{eqnarray}
Eq. (\ref{J}) simplifies to 
\begin{eqnarray}
\frac{4(\tilde{u}+1)^3(\tilde{u}-1)}{4\tilde{u}+5}=\frac{E_4^3}{E_6^2-E_4^3}\ ,
\end{eqnarray}
where $\tilde{u}=\frac{2^{\frac{2}{3}}u}{3}$. The equation for $u$
has 4 solutions and 3 of them have the correct asymptotic
behavior $u\rightarrow \infty$ in the weak coupling limit
$\tau\rightarrow i\infty$. These 3 solutions are related by a
$Z_3$ transformation and we just need to consider one
solution
\begin{eqnarray}
\label{uoftauconformal}
u=\frac{3}{2^{\frac{5}{3}}}\big{(}-1+\sqrt{x}+\sqrt{3-x+\frac{2E_6^2}{(E_6^2-E_4^3)\sqrt{x}}}~\big{)}
\end{eqnarray}
with  $x=1+\frac{E_4^2}{(E_6^2-E_4^3)^{\frac{2}{3}}}-\frac{E_4}{(E_6^2-E_4^3)^{\frac{1}{3}}}$.

The formula (\ref{aumassiveNf1}) becomes 
\begin{eqnarray}
\frac{d a}{d
u}=\big{(}\frac{2^{\frac{2}{3}}(\tilde{u}+1)}{6(2\tilde{u}^2+2\tilde{u}-1)}\frac{E_6}{E_4}\big{)}^{\frac{1}{2}}
\end{eqnarray}
and the genus one amplitude becomes
\begin{eqnarray}
F^{(1)} &=&
-\frac{1}{2}\textrm{log}(\frac{da}{du})-\frac{1}{12}\textrm{log}(\Delta)
\nonumber \\
&=&
-\frac{1}{12}\textrm{log}(\frac{(4\tilde{u}+5)(\tilde{u}+1)^3(\tilde{u}-1)^2}{(2\tilde{u}^2+2\tilde{u}-1)^3}\frac{E_6^3}{E_4^3})\ ,
\end{eqnarray}
which can be written entirely in terms of $\tau$ using (\ref{uoftauconformal}).

We study the gravitational couplings of the theory at the superconformal point 
$u=\frac{3}{2^{\frac{2}{3}}}$ in more details. The Picard-Fuchs equation has a
constant solution around this point and two power series solutions. The two 
power series serve as the flat coordinate $a$ and the derivative of prepotential $\frac{\partial F^{(0)}}{\partial a}$  around this point. Denoting $z=u-u_1\rightarrow 0$, we find the solutions are 
\begin{eqnarray}
a &=& z^{\frac{5}{6}}(1-\frac{10}{891}2^{\frac{2}{3}}z+\frac{128}{111537}2^{\frac{1}{3}}z^2-\frac{6272}{36669429}z^3+\mathcal{O}(z^4)), \nonumber \\
\frac{\partial F^{(0)}}{\partial a} &=& z^{\frac{7}{6}}(1-\frac{28}{1053}2^{\frac{2}{3}}z+\frac{400}{124659}2^{\frac{1}{3}}z^2 
-\frac{4096}{7971615}z^3+\mathcal{O}(z^4))\ .
\end{eqnarray}
The scaling behavior of the period of the Picard-Fuchs equation as $a\sim
z^{\frac{5}{6}}$ agrees with the analysis 
presented in~\cite{Argyres}. We can invert the series and solve the prepotential in terms of the flat coordinate
\begin{eqnarray} 
F^{(0)} = a^{\frac{12}{5}}-\frac{28}{3861}2^{\frac{2}{3}}a^{\frac{18}{5}}+\frac{63872}{142457535}2^{\frac{1}{3}}a^{\frac{18}{5}}
-\frac{11006912}{256086163575}a^6 +\mathcal{O}(a^{\frac{36}{5}})\ . 
\end{eqnarray}
Here we have not fix the normalization for prepotential and the flat
coordinate $a$. As usual the prepotential is determined up to a quadratic polynomial of $a$ due to the ambiguity in choosing the basis of Picard-Fuchs equation. 

For the genus one free energy, we find 
\begin{eqnarray}
F^{(1)}=-\frac{1}{10}\log(a)-\frac{1}{891}2^{\frac{2}{3}}a^{\frac{6}{5}}+\frac{5732}{22493295}2^{\frac{1}{3}}a^{\frac{12}{5}}
+\mathcal{O}(a^{\frac{18}{5}}) \ .
\end{eqnarray}

For genus two free energy, we can use the result we derive for generic mass parameter $m_1$, and specialize to the superconformal point. Using formulae (\ref{3.77-10-31-2008}) we can derive the expression for the Eisenstein series
\begin{eqnarray}
&&E_4(\tau) = 12g_2 (\frac{d a}{d u})^4, ~~~~~~
E_6(\tau) = -216 g_3 (\frac{d a}{d u})^6, \nonumber \\
&& X(\tau) = \frac{E_2(\tau)E_4(\tau)}{E_6(\tau)}= 
\frac{2E_4(\tau)E_6(\tau)-3E_4(\tau)^3  (\frac{dE_4(\tau)}{d \tau})/(\frac{dE_6(\tau)}{d \tau})}
{2E_4(\tau)E_6(\tau)-3E_6(\tau)^2  (\frac{dE_4(\tau)}{d
    \tau})/(\frac{dE_6(\tau)}{d \tau})}\ .
\end{eqnarray}
We find that the leading singularity at genus two is $F^{(2)}=\mathcal{O}(\frac{1}{z})$, even though for a generic holomorphic 
ambiguity $f^{(2)}$ one has a leading singular behavior  of $f^{(2)}=\mathcal{O}(\frac{1}{z^4})$. The expansion in flat coordinate is 
\begin{eqnarray}
F^{(2)} = \frac{133}{38880 a^{\frac{6}{5}}}\! +\! \frac{19}{360855\cdot
  2^{\frac{1}{3}}}\! -\! 
\frac{79793}{1656324450}\frac{a^{\frac{6}{5}}}{2^{\frac{2}{3}}}\! +\! 
\frac{4310932}{332775068175}a^{\frac{12}{5}}\! +\! \mathcal{O}(a^{\frac{18}{5}})\ .
\end{eqnarray}
This kind of singularity behavior is very similar to the situation at
the orbifold singularity in compact Calabi-Yau spaces encountered in
\cite{HKQ}, where the $F^{(g)}$ turns out to be less singular than 
naively expected. Here the massless particles scale as the period $a\sim z^{\frac{5}{6}}$ in the
limit $z\rightarrow 0$ and we know $F^{(g)}$ should be no more singular 
than $a^{2-2g}$ from the usual Gopakumar-Vafa argument of integrating out charged 
particles in the graviphoton background.  This explains the leading 
singularity in the expression of $F^{(2)}$ above.

\subsection{$N_f=2$, $m_1=m_2=\pm\frac{\Lambda_2}{2}$}
We follow our previous notation $\Lambda_2=2$. There is a $(2,1)$
superconformal field theory at $u=\frac{3}{2}$ where a double
singularity collides with a mutually non-local dyon singularity,
and there is another dyon singularity at $u=-\frac{5}{2}$. The
discriminant is
\begin{eqnarray}
\Delta\sim(u-\frac{3}{2})^3(u+\frac{5}{2})\ .
\end{eqnarray}
We solve a cubic equation for $u$, and there are 2 solutions with
the correct asymptotic behavior in weak coupling limit. As before
they are related by a $Z_2$ symmetry. We consider one solution
\begin{eqnarray} \label{superconformal(2,1)u}
u&=&
-\frac{3}{2}+\frac{3e^{-\frac{\pi}{6}i}E_4(E_6+i(E_4^3-E_6^2)^{\frac{1}{2}})^{\frac{1}{3}}}{2(E_4^3-E_6^2)^{\frac{1}{2}}}
\nonumber \\ &&
+\frac{3e^{\frac{\pi}{6}i}E_4(E_6-i(E_4^3-E_6^2)^{\frac{1}{2}})^{\frac{1}{3}}}{2(E_4^3-E_6^2)^{\frac{1}{2}}}\ . 
\end{eqnarray}
From (\ref{aumassiveNf1}) one gets
\begin{eqnarray}
\frac{d a}{d u}=
\frac{1}{2}\big{(}\frac{(2u+3)}{(2u-3)(u+3)}\frac{E_6(\tau)}{E_4(\tau)}\big{)}^{\frac{1}{2}} \ , 
\end{eqnarray}
which allows to write the genus one amplitude as
\begin{eqnarray}
F^{(1)} &=&
-\frac{1}{2}\textrm{log}(\frac{da}{du})-\frac{1}{12}\textrm{log}(\Delta)
\nonumber \\
&=&
-\frac{1}{12}\textrm{log}(\frac{(2u+3)^3(2u+5)}{(u+3)^3}\frac{E_6^3}{E_4^3})\ .
\end{eqnarray}
One can use the expression of $u$ in equation
(\ref{superconformal(2,1)u}) to obtain a formula for genus one
amplitude $F^{(1)}$ purely in terms of gauge coupling $\tau$.

There are two other identical $(2,1)$ superconformal field theories
at $m_1=-m_2=\pm i\frac{\Lambda_2}{2}$. These are the same theories 
as the ones at $m_1=m_2=\pm\frac{\Lambda_2}{2}$, to which they are 
related by the transformation $u\rightarrow -u$.

As in the $N_f=1$ case, we can solve the flat coordinate $a$ and express the prepotential $F^{(0)}$ and genus one free energy $F^{(1)}$ in terms of it. The best way to solve the flat coordinate is to use the Picard-Fuchs equation, since at the superconformal point it is not convenient to do perturbative expansion the Eisenstein series. The Picard-Fuchs equation for the massive $N_f=2,3$ Seiberg-Witten theory were found in \cite{Ohta2}. We specialize to the case of mass $m_1=m_2=\pm\frac{\Lambda_2}{2}$. Denote $z=u-\frac{3}{2}$ we found the solutions of Picard-Fuchs equation around $z=0$ as the followings 
\begin{eqnarray}
 a &=&  z^{\frac{3}{4}}(1-\frac{3}{224}z+\frac{25}{22528}z^2-\frac{9}{65536}z^3+\mathcal{O}(z^4)), \nonumber \\
 \frac{\partial F^{(0)}}{\partial a} &=& z^{\frac{5}{4}}(1-\frac{5}{96}z+\frac{147}{26624}z^2-\frac{847}{1114112}z^3+\mathcal{O}(z^4)) \ . 
\end{eqnarray}
The genus zero and one free energy up to a constant are 
\begin{eqnarray}
F^{(0)} &=& a^{\frac{8}{3}}-\frac{5}{252}a^4+\frac{3197}{2690688}a^{\frac{16}{3}}-\frac{6883}{65680384}a^{\frac{20}{3}}+\mathcal{O}(a^8) \nonumber \\
F^{(1)} &=& -\frac{1}{6} \log (a) -\frac{5}{672} a^{\frac{4}{3}}+\frac{1409}{1655808}a^{\frac{8}{3}}-\frac{11873}{92725248}a^4+\mathcal{O}(a^{\frac{16}{3}})\ .
\end{eqnarray}

\subsection{$N_f=3$, $m_1=m_2=m_3=\frac{\Lambda_3}{8}$}
We follow our previous notation $\Lambda_3=4$. In this case there
is a $(3,1)$ superconformal field theory at $u=\frac{1}{2}$ where
a triple singularity collides with a dyon singularity, and there is
also a dyon point at $u=-\frac{19}{16}$. The discriminant is
\begin{eqnarray}
\Delta\sim(u-\frac{1}{2})^4(u+\frac{19}{16})\ .
\end{eqnarray}
One can solve an algebraic equation and obtain an expression of $u$
in terms of gauge coupling $\tau$. Here we will use the
normalization
$\tau=\frac{i}{2\pi}\frac{\partial^2F^{(0)}}{\partial a^2}$, which
is one half of the T-dual of the gauge coupling we use in $N_f=3$
massless case previously in Section \ref{sectionmassless}. There
are two branches of solutions~\cite{Brandhuber}, and we take the
branch where in the weak coupling limit $\tau\rightarrow i\infty$,
the modulus goes like $u\rightarrow \infty$. The expression for
$u$ is \cite{Brandhuber}
\begin{eqnarray}
u=\frac{27E_6E_4^{\frac{3}{2}}+23E_4^3+4E_6^2}{8(E_6^2-E_4^3)}
\end{eqnarray}
and there are also formulae for the derivative of period $a$
\begin{eqnarray}
\frac{da}{du}&=&(\frac{2}{27})^{\frac{1}{2}}(\frac{E_6-E_4^{\frac{3}{2}}}{E_4})^{\frac{1}{2}}
\nonumber \\
\frac{da}{d\tau} &=&
-(\frac{27}{128})^{\frac{1}{2}}\frac{E_6+E_4^{\frac{3}{2}}}{(E_6-E_4^{\frac{3}{2}})^{\frac{1}{2}}}\ .
\nonumber
\end{eqnarray}
We have not found an explicit formula for period $a(\tau)$.
However, to write an exact formula for the topological string
amplitude of the $F^{(g)}$ in terms of modular forms, we only need the
derivative of period $a$. The genus one amplitude is
\begin{eqnarray}
F^{(1)} &=&
-\frac{1}{2}\textrm{log}(\frac{da}{du})-\frac{1}{12}\textrm{log}(\Delta)
\nonumber \\
&=&
-\frac{1}{12}\textrm{log}(\frac{E_4^3(E_6+E_4^{\frac{3}{2}})}{(E_6-E_4^{\frac{3}{2}})^2})
\end{eqnarray}
where as usual we are not careful about an ambiguous additive
constant in $F^{(1)}$.

As in the $N_f=1,2$ case, we can solve the flat coordinate $a$ and express the prepotential $F^{(0)}$ and genus one free energy $F^{(1)}$ in terms of it.  Denote $z=u-\frac{1}{2}$, we found the solutions of Picard-Fuchs equation around $z=0$ as the followings 
\begin{eqnarray}
 a &=&  z^{\frac{2}{3}}(1-\frac{8}{405}z+\frac{49}{13122}z^2-\frac{18928}{17537553}z^3+\mathcal{O}(z^4)), \nonumber \\
 \frac{\partial F^{(0)}}{\partial a} &=& z^{\frac{4}{3}}(1-\frac{80}{567}z+\frac{242}{6561}z^2-\frac{254320}{20726199}z^3+\mathcal{O}(z^4))\ .  
\end{eqnarray}
The genus zero and one free energy up to a constant are 
\begin{eqnarray}
F^{(0)} &=& a^3-\frac{64}{945}a^{\frac{9}{2}}+\frac{401}{36450}a^6-\frac{885232}{351833625}a^{\frac{15}{2}}+\mathcal{O}(a^9) \nonumber \\
F^{(1)} &=& -\frac{1}{4} \log (a) -\frac{4}{135} a^{\frac{3}{2}}+\frac{3403}{437400}a^{3}-\frac{225869}{81192375}a^{\frac{9}{2}}+\mathcal{O}(a^{6})\ .
\end{eqnarray}

\section{The non-compact limit of Calabi-Yau compactifications}
\label{locallimit} 
A good way to solve the holomorphic anomaly equations for the Riemann surface 
is in terms of rings of an-holomorphic modular forms, as we have seen in 
sections 2 and 3. 

However even without knowing anything about the structure of the
modular forms w.r.t. the  modular group of the family of curves, one can derive 
the necessary almost holomorphic objects directly in terms of the 
periods, which are  solutions of the Picard-Fuchs equations.
This has been done for compact Calabi-Yau spaces
using special K\"ahler geometry~\cite{BCOV} and the anholomorphic objects are
the BCOV propagators. The derivatives that appear in the holomorphic anomaly
equation close within a finitely generated polynomial ring of almost 
holomorphic modular functions and the $F^{(g)}$ are themselves such 
polynomials~\cite{Yamaguchi}. The ring structure has been further analyzed 
in~\cite{Grimm:2007tm,Alim:2007qj,Hosono}. 

As explained in~\cite{Kachru:1995fv,Katz:1996fh} extracting  4d  $N=2$ non-perturbative gauge 
theory from type II string theory compactified on a Calabi-Yau space $W$ can be done 
geometrically by taking a limit in the geometrical parameter of the Calabi-Yau space 
in which part of the geometry decompactifies. Since we are dealing with the vector moduli space 
the limit is taken in the A-model in the complexified K\"ahler space and in the B-model 
on the mirror manifold the limit is is taken in the complex structure space. 
For Calabi-Yau manifolds $W$ embedded in toric ambient spaces a wide class of limiting configurations in 
the B-model has be described as the 3-d total space of a conic bundle over 
$\mathbb{C}^*\times \mathbb{C}^*$ branched at a Riemann surface 
${\cal C}^*_g\in \mathbb{C}^*\times \mathbb{C}^*$~\cite{Katz:1996fh}. 
For the relevant geometries the compactification of ${\cal C}^*_g$ is 
then the Seiberg-Witten curve ${\cal C}_g$. A general feature of the limit is that 
the periods of the $(3,0)$-form $\Omega$ over 3-cycles in $W$, which stay finite become 
the periods of a meromorphic form $\lambda$ over 1-cycles on ${\cal C}^*_g$~\cite{Kachru:1995fv}. 
The form $\lambda$ on ${\cal C}^*_g$ can be obtained by integrating 
$\Omega$ over the non-compact directions in the limiting configuration 
of $W$~\cite{Klemm:1996bj}.

The non-compact limit relates the Calabi-Yau rings of~\cite{BCOV,Yamaguchi,Grimm:2007tm,Alim:2007qj,Hosono} 
to the classical rings of almost holomorphic modular forms of subgroups 
of $SL(2,\mathbb{Z})$ for $SU(2)$ gauge groups~\cite{SW1} or $SP(2g,\mathbb{Z})$ for 
$SU(g+1,\mathbb{Z})$ gauge groups\footnote{Seiberg-Witten curves 
are by now known for any gauge group. In general they are special families
of complex curves, whose deformation parameters correspond to vev of fields in 
the Cartan subalgebra of the group, see~\cite{Klemm:1997gg} for a  
review.}~\cite{Huang,Aganagic:2006wq} and it  must be  possible 
to write the generators of the an-holomorphic rings that are needed to 
solve the holomorphic anomaly equation for gauge theories in terms 
of solutions of the Picard-Fuchs equations.  For example for $SU(3)$ 
in terms of the solutions for the Appel differential system~\cite{KLT}. 
Below we discuss the properties of the limit and how the ring structure
behaves in the non-compact limit, extending the work 
of~\cite{Haghighat:2008gw,Alim:2008kp}.

\subsection{Special geometry and rigid special geometry} 
\label{specialgeometry} 
The origin of the an-holomorphicity comes from the metric on the moduli 
space of the $N=2$ vector multiplets, which determines their kinetic term
in the effective action. The latter is an $N=2$ supergravity action 
for the compact case and an $N=2$ super symmetric (gauge theory) action  
without gravity in the non-compact case. The vector multiplet  moduli 
spaces are identified  with the complex structure moduli space ${\cal M}$ of the
Calabi-Yau $W$ and the Riemann-surface ${\cal C}^*_g$ respectively. The
metrics are the Weil-Petersen metrics on these geometric moduli spaces. 
In both cases they derive from a real K\"ahler potential $K$ as
$G_{i\bar\jmath}=
\partial_i\bar \partial_{\bar \jmath} K$, but there is 
additional structure. For the Calabi-Yau case this is usually formulated 
as special K\"ahler geometry in the small phase space, i.e. in the 
inhomogeneous coordinates discussed below, and for the Riemann-surface the structure 
is always rigid special geometry. However in the homogeneous 
coordinates, also called the big moduli space,  the $N=2$ supergravity action for 
the compact case can also be formulated in terms of rigid special geometry, 
which simplifies the limit to the local case.

The splitting of the middle de Rham cohomology of the compact Calabi-Yau $W$
\begin{equation}
\begin{array}{rl}
H^{3}(W,\mathbb{Z})=&H^{3,0}\oplus H^{2,1}\oplus H^{1,2}\oplus H^{0,3}\\ 
& \Omega\qquad \  \chi_i\qquad \,  \bar \chi_{\bar\imath} \qquad\    \bar \Omega,\qquad\qquad   i,\bar \imath=1,\ldots,h^{21}
\end{array}
\label{hodge}
\end{equation} 
into Hodge cohomology groups depends on the choice of complex structure. 
In particular we indicated the basis  $\Omega, \chi_i$, $\bar \chi_{\bar\imath}$, $\bar \Omega$, 
$i,\bar \imath=1,\ldots,h^{21}$  of the individual Hodge cohomology groups that changes 
with the complex structure.
One introduces a fixed topological and symplectic basis $(A^K,B_K)$ of $H_3(W,\mathbb{Z})$ 
and a dual symplectic basis  $(\alpha_K,\beta^K)$ of $H^{3}(W,\mathbb{Z})$. 
Here $K=0,\ldots,h_{21}$ and  the non vanishing pairings are 
$A^L\cap B_K=\int_W \alpha_K\wedge \beta^L:=\langle \alpha_K,\beta^L\rangle=-\langle \beta^L, \alpha_K\rangle=\int_{A^L} 
\alpha_K=\int_{B_K} \beta^L=\delta^L_K$. If one expand 
$\Omega=X^I \alpha_I-F_I \beta^I$ and $\chi_i=\chi^I_i\alpha_I-\chi_{Ii}\beta^I$ in terms 
of periods 
\begin{equation} 
X^I=\int_{A^I} \Omega,\qquad F_i=\int_{B_I} \Omega, \qquad \chi^I_i=\int_{A^I}
\chi_i, \qquad \chi_{Ii}=\int_{B_I} \chi_i \ , 
\end{equation} 
the $X^I$ become homogeneous coordinates of the moduli space of complex structures. 
The dual periods $F_I=\int_{B_I} \Omega$ and the $\chi^I_i,\chi_{Ii}$ are not independent but 
related to $X^I$ by special geometry. 
It is convenient to define $Y^I:=\lambda^{-1} X^I$, $I=0,\ldots,h_{21}$ and $\partial_I=\partial_{Y_I}$.
It is easy to see that $\partial_{I_1},\ldots,\partial_{I_k} \Omega \in \bigoplus_{l=0,k} H^{3-l,l}$. 
Transversality, i.e. $\langle a,b\rangle:=\int_W a\wedge b=0$ unless  both Hodge indices $(p,q)$ of $a$ and $b$ 
add up to $3$, means that $\langle \Omega, \partial_I\Omega\rangle= \langle \Omega, \partial_I\partial_J\Omega\rangle =0$ and 
that implies the existence of a holomorphic prepotential $F^{(0)}({\underline Y})=\frac{1}{2} \lambda^2 Y^I F_I$. The latter 
is a homogeneous function of degree two in $Y^I$, such that $F_I=\frac{\partial F^{(0)}}{\partial Y^I}$.
In the big moduli space, parametrized by the homogeneous coordinates $Y^I$, one defines a K\"ahler potential
\begin{equation}
{\cal  K}=\frac{i}{2} (Y^K \bar F_K- \bar Y^K F_K), \qquad {\cal G}_{IJ}=\partial_I\partial_J {\cal K}={\rm Im}\, \tau_{IJ}\ ,
\label{rigidkaehler}
\end{equation}  
a symmetric weight zero tensor $\tau_{IJ}=\frac{\partial^2 F}{\partial Y_I \partial Y_J}$ and a symmetric 
weight $-1$ triple coupling $C_{IJK}=\partial_I \partial_J\partial_K F=
\langle \Omega, \partial_I\partial_J \partial_K \Omega \rangle$. The metric  ${\cal G}_{IJ}$ has signature $(1,h_{21})$.
The connection is $\Gamma_{IJ}^K={\cal G}^{K\bar L}\partial_J {\cal G}_{I\bar L}=-\frac{i}{2} C_{JK}^I$ 
and one has the so called special geometry relation 
\begin{equation}
[D_{\bar I},D_J]_K^L=\bar \partial_{\bar I} \Gamma_{JK}^L =\frac{1}{4}C_{JKP}\bar C_{\bar I}^{PL} \ ,
\label{specialgeometrybigphasespace}
\end{equation}
which can be viewed as integrability condition for the existence of the holomorphic prepotential $F$, such that $C_{IJK}$ 
and ${\cal G}_{I\bar J}$ can be obtained from it by differentiation.   
It was shown in~\cite{Grimm:2007tm} that the holomorphic anomaly equation of~\cite{BCOV} reads in the big moduli space 
\begin{equation} 
\bar \partial_{\bar I} F^{(g)}=-\frac{i}{8} \bar C^{JK}_{\bar I}
\left(D_J\partial_K F^{(g-1)}+\sum_{h=1}^{g-1} \partial_J F^{(h)} \partial_K F^{(g-h)}\right)\ .
\label{anomalybigphasespace}
\end{equation} 
Since $\bar D_{\bar I} \bar C_{\bar J\bar K\bar L}=\bar D_{\bar J} \bar C_{\bar I\bar J \bar J}$ and 
$\bar D_{\bar K} {\cal G}_{I\bar L}=0$ one can integrate 
\begin{equation}
\label{defbigprop} 
\partial_{\bar K} {S}^{IJ}=\frac{i}{4}\bar C^{IJ}_{\bar K}\ 
\end{equation} 
The $S^{IJ}$ are an-holomorphic tensors, called  the propagators\footnote{Indeed the $F^{(g)}$ 
can be reconstructed with the propagators and vertices $\partial_{I_1},\ldots, \partial_{I_n} F^{(h<g)}$ 
by Feynman rules of an auxiliary field theory~\cite{BCOV} for the small phase space. For the 
formalism in the big phase space see~\cite{Grimm:2007tm}.}, they play a similar r\^ole that $\hat E_2$ plays 
for the elliptic curve.
{}From (\ref{specialgeometrybigphasespace}) one can solve 
\begin{equation}
\label{bigprop} 
S^{KL}=i (C_I^{-1})^{KP} (\Gamma_{I})_P^L+{\cal H}^{KL} \ .
\end{equation}
Here $I$ is not summed over and ${\cal H}^{KL}$ is a holomorphic ambiguity. The latter must be chosen so that 
$S^{KL}$ is a tensor. The precise choice affects the form of the total ambiguity $f^{(g)}$, but is otherwise arbitrary. One 
convenient choice is to require that $\partial_I F^{(1)}=-\frac{i}{8}C_{IKL} S^{KL}$.

The idea of direct integration is based on the fact that all anholomorphic
dependence of the modular invariant scalar $F^{(g)}$ is in the $S^{IJ}$. 
Therefore by (\ref{defbigprop}) $\bar \partial_{\bar I} F^{(g)}=\frac{i}{4} 
\bar C_{\bar I}^{JK} \frac{\partial F}{\partial S^{JK}}$ and $\bar C^{JK}_{\bar I}$ cancels from equation 
(\ref{anomalybigphasespace}), which can then be {\em directly integrated}  w.r.t. to $S^{JK}$ 
up to holomorphic terms $f^{(g)}$, which must also be modular invariant. To proceed  in the iteration in $g$
one must show that the anholomorphic generators $S^{IJ}$ close under the covariant derivative up to holomorphic 
terms. By considering $\bar \partial_{\bar I} D_I S^{JK}$, using 
(\ref{specialgeometrybigphasespace}) and integrating w.r.t $\bar Y^{\bar I}$ one gets
\begin{equation}
\label{closing}
D_I S^{JK}=C_{IMN} S^{NK} S^{MJ}+{\cal H}_I^{JK}\ .
\end{equation}

Let us now come to special geometry in the small phase space, whose
coordinates are the inhomogeneous variables $t^i=\frac{X^i}{X^0}$, $i=1,\ldots, h_{21}$. 
The K\"ahler potential $K$ in the small phase space is given by\footnote{We follow the 
conventions of \cite{Grimm:2007tm}.} 
\begin{equation} 
e^{-K}=i\int_W \Omega\wedge \bar \Omega=i (X^I {\bar F}_{\bar I}-{\bar
  X}^{\bar I} F_I)=i(t^i-{\bar t}^{\bar \imath})(\partial_i {\cal F}^{(0)}+{\bar \partial}_{\bar \imath}
{\overline {\cal F}}^{(0)})-2i({\cal F}^{(0)}-{\overline {\cal F}}^{(0)})\ . 
\label{kaehler}
\end{equation}
Here  we define $(X^0)^2{\cal F}^{(0)}({\underline t})=F^{(0)}({\underline X})$ 
using the degree 2 homogeneity of  $F^{(0)}$ and the third equality holds up
to a K\"ahler transformation.

The connection $\Gamma^{I}_{JK}$ splits into a metric connection, w.r.t.
$G_{i\bar \jmath}=\partial_i \partial_{\bar \jmath} K$, and 
a K\"ahler connection. The covariant derivative becomes 
$D_i=\partial_j-\Gamma_i-k K_i$ for objects in  
${\cal L}^k\otimes T^*{\cal M}$, with an analogous definition 
for $D_{\bar \imath}$. Holomorphic sections $A({\underline t})$ of ${\cal L}^k$ 
transform like $A({\underline t})\rightarrow A({\underline t}) e^{-k h({\underline t})}$ 
under K\"ahler transformations $K({\underline t},\bar {\underline t})\rightarrow K({\underline t}, 
\bar {\underline t})+h({\underline t}) +\bar h(\bar {\underline t})$.  
In particular the holomorphic $(3,0)$ form $\Omega\in {\cal L}$ and $F^{(g)}\in {\cal L}^{2-2g}$. 
The covariant derivative eliminates the $(3,0)$ part in the derivative of $\Omega$ and hence 
$\chi_i=D_i\Omega$ ($\bar \chi_{\bar\imath}=D_{\bar \imath} \bar \Omega$). Applying this 
under the integral yields $\chi_i^I=D_iX^I$ and $\chi_{Ii}=D_iF_I$, which serve as 
projectors from the big to the small phase space. In particular the triple coupling 
in inhomogeneous variables $C_{ijk}\in {\cal L}^2\otimes {\rm Sym}^3 T^*{\cal M}$ are
\begin{equation}
C_{ijk}=\langle \Omega,\partial_i\partial_j\partial_k\Omega\rangle=
\sum_{I=0}^{h_{21}}(X^I\partial_i\partial_j\partial_k F_I-
F_I\partial_i\partial_j\partial_k X^I)=\chi_i^I\chi_j^J\chi_k^K C_{IJK}\ .
\label{complextriplecoupling3}
\end{equation}
It follows that $C_{ijk}=D_i D_j D_k {\cal F}^{(0)}({\underline t})$.  
Using  $\langle \chi_i,\bar \chi_{\bar\imath}\rangle=e^{-K} G_{i\bar \imath}$ 
from (\ref{kaehler}) and transversality one gets with the 
definition (\ref{complextriplecoupling3}) 
\begin{equation}  
D_iX^I=:\chi^I_i, \qquad D_i\chi_j^I=iC_{ijk} G^{k\bar k}\bar \chi_{\bar \jmath}^I e^K, 
\qquad D_i \bar \chi_{\bar \imath}^I=G_{i\bar \imath}\bar X^I\ .
\end{equation}
With $[D_i,\bar D_{\bar \imath}] \chi_k=-G_{i\bar \imath}\chi_k
+ R_{i\bar \imath k}^{\phantom{ijk}l}\chi_l$ one arrives at the special K\"ahler relation 
in inhomogeneous coordinates
\begin{equation} 
[D_i,D_{\bar \imath}]_j^k=R_{i \bar \imath j}^{{\phantom{iij}}k}= {\bar \partial}_{\bar \imath} \Gamma^k_{ij}
=\delta_{i}^k G_{j\bar \imath}+\delta_{j}^k G_{i\bar \imath}- C_{ijl} C^{kl}_{\bar \imath}\ \ .
\label{sk} 
\end{equation}
The projection of the $S^{IJ}$ is straightforward
\begin{equation} 
S^{IJ}=(X^I\  \chi_i^I)\left(\begin{array}{cc} S&-S^i
\\ -S^i& S^{ij}\end{array}\right) \left(\begin{array}{c} X^J
\\ \chi_j^J\end{array}\right)\ .
\end{equation} 
Here the relations 
$\bar D_{\bar \imath} \bar C_{\bar \jmath \bar k \bar l}=\bar D_{\bar \jmath} \bar C_{\bar 
\imath \bar k \bar l}$ are integrated to $\bar C^{jk}_{\bar \imath}=  \bar \partial_{\bar \imath} S^{jk}$, 
$G_{\bar \imath k} S^{kj}= 
\bar \partial_{\bar \imath } S^j$ and $G_{\bar \imath j} S^{j} = \frac{1}{2} \bar \partial_{\bar \imath}S $.
The potentials $S^{ij},S^j,S$, also called the propagators, allow  
to solve the anomaly equation, by partial integration, see for 
details~\cite{BCOV}, up to an holomorphic ambiguity. One can project the
propagators from the big phase space or rederive them from the projected
special K\"ahler relation (\ref{sk}). E.g. $S^{ij}$ is solved from  (\ref{sk})  
\begin{equation} \label{e:propeq} 
\Gamma^k_{ij}=\delta_{i}^k \partial_j K+\delta_{j}^k \partial_i K- C_{ijl}
S^{kl}+\tilde h_{ij}^k\ . 
\end{equation}                  
The analogs of the statement about the closing of the propagators (\ref{closing})
under $D_i$ are~\cite{Yamaguchi,Alim:2007qj}
\begin{equation}
\begin{array}{rl}
D_i S^{kl} &=  \delta_{i}^k S^l+\delta_{i}^l S^k - C_{inm} S^{km} S^{ln}+ h^{kl}_i\ ,\\[ 2mm]
D_i S^j    &= 2 \delta_i^j- C_{ikl} S^{kl} + h_i^{jk} K_k+h^j_i\ , \\[2mm]   
D_i S      &= C_{ikl}  S^k S^l+\frac{1}{2} h^{kl}_i K_k K_l+ h_i^l K_l+ h_i\ , \\[2 mm] 
D_i K_j    &=- K_i K_j -C_{ijk} S^k+ C_{ijk} S^{kl} K_l+ h_{ij}\  
\end{array}
\label{closingsmall} 
\end{equation}
and are derived from special geometry similarly as (\ref{closing}). 
E.g. from ${\bar \partial}_{\bar k}(D_i S^{kl}) ={\bar \partial}_{\bar k} 
(  \delta_{i}^kS^l+\delta_{i}^l S^k - C_{inm} S^{km} S^{ln})$ follows the
first of the closing relations (\ref{closingsmall}), etc.

One finds from the properties under K\"ahler transformations~\cite{BCOV,Yamaguchi,Alim:2007qj}~that  
$\tilde S^{ij}=S^{ij},\tilde S^j=S^i-S^{ij} K_j$ and $\tilde S=S-S^iK_i+\frac{1}{2}S^{ij}K_i K_i$ 
are a complete set of an-holomorphic generators of a polynomial 
ring that contains the $F^{(g)}$ as polynomials with holomorphic 
coefficients. Indeed one can write the holomorphic anomaly equation as 
\begin{equation} 
\frac{\partial {\cal F}^{(g)}}{\partial S^{ij}}=\frac{1}{2} \left(D_i D_j 
{\cal F}^{(g-1)} + \sum_{h=1}^{g-1} D_i {\cal F}^{(g-h)} D_j {\cal F}^{(h)}\right)
\end{equation} 
and integrate if up to holomorphic terms as a polynomial. Note that the 
derivatives of ${\cal F}^{(g)}$ w.r.t. $S^j,S$ and $K_i$, which naively occur at the 
left hand side, cancel. This cancellation is equivalent to the statement that the 
dependence of $F^{(g)}$ is through the combinations $\tilde S^{ij},\tilde S^j$ and $\tilde S$.

\subsection{The non-compact limit}

Non-compact Calabi-Yau are mirror to 
\begin{equation} 
uv =H(x,y;\underline{ z}),
\label{noncompactmirror}
\end{equation}
where $u,v\in \mathbb{Z}$, $x,y \in \mathbb{C}^*$ and $\underline {z}$ are 
moduli of the geometry.  The geometry is that of conic bundle, which branches 
over the locus  
\begin{equation} 
H(x,y;\underline{z})=0,
\end{equation}
which is  a family Riemann surfaces ${\cal C}^*_g$ of genus $g$.  
Let  $\lambda=\log(x) \frac{\dd y}{y}$ be meromorphic 
differential and $(a^i,b_i)$ a symplectic basis of 
$H^1(\Sigma_g,\mathbb{Z})$ then  the  rigid effective action has a K\"ahler 
potential~\footnote{We use here conventions, which differ by a factor $i$ multiplying the prepotential 
from the ones used in (\ref{metric}) and call the flat coordinates $t^i,{\cal F}_i$ instead of  $a^i,a_{D_i}$.}
\begin{equation}
K=\frac{i}{2} (t^i {\bar {\cal F}}_{\bar \imath }- {\bar t}^{\bar \imath}   {\cal F}_i)\ ,
\label{localKaehler}
\end{equation}
where $t^i=\int_{a^i} \lambda$ and ${\cal F}_i=\int_{b_i} \lambda$. Note 
that the form of $K$ is like in (\ref{rigidkaehler}), but the $t^i$  are directly 
appropriate flat local coordinates. The metric reads
\begin{equation}
G_{i\bar \jmath}=\partial_i{\bar \partial}_{\bar \jmath} K=\frac{1}{2 i}
\left(\tau_{ij}- \bar \tau_{\bar \imath \bar \jmath}\right),
\end{equation}         
where $\tau_{ij}=\frac{\partial^2 {\cal F}}{\partial t^i\partial t^j}$.

In the local case one has the following simplifications. 
The K\"ahler connection in $D_i$ becomes trivial, and the $S^l$ as well as the $S$ (see \cite{Klemm:1999gm}) 
vanish, \emph{i.e.} the first equation in (\ref{closingsmall}) and the equation (\ref{closing}) become equivalent 
and read
\begin{equation} \label{DS}
D_i S^{kl} = -C_{inm} S^{km} S^{ln}+ f^{kl}_i .
\end{equation} 

The  $S^{ij}$ are the generators of the ring of anholomorphic objects
Since the K\"ahler connection $\partial_j K$ in (\ref{e:propeq}) drops out, so
the $S^{ji}$ are solved from 
\begin{equation} \label{PropEq}
\Gamma^k_{ij}=- C_{ijl} S^{kl}+\tilde f_{ij}^k\ 
\end{equation}  
as 
\begin{equation} 
S^{ij}=-(C_p)^{il} \left[(\Gamma_p)_{l}^j+({\tilde f}_p)^j_l\right],\quad \forall \ p=1,\ldots, r\ .
\label{localprop}
\end{equation} 
Here $r$ is the number of K\"ahler parameter in the mirror to~\ref{noncompactmirror}. 
It has been pointed out e.g. in~\cite{Haghighat:2008gw} that there are in general 
algebraic relations between the $S^{ij}$. If ${\cal C}_f$ has genus $g=1$ there will
be only one independent $S^{ij}$, for $g=2$ there should be $3$ independent $S^{ij}$. 
Again $p$ is not summed over in~(\ref{localprop}) and this over determined system 
requires a suitable choice  of the ambiguity $\tilde f_{ij}^k$. This choice is 
simplified by the fact~\cite{Aganagic:2002wv} that $\d_i F_1$ can be expressed through the 
propagator as
\begin{equation}
\partial_i F_1=\half C_{ijk}S^{jk}+A_i,
\label{F1prop}
\end{equation}
with an ambiguity $A_i$, which can be determined by the ansatz $A_i=\d_i(a\log\Delta+b_j\log z_j)$.  
Moreover the universal behavior of $F_1$ near the conifold 
locus~\cite{Ghoshal:1995wm} implies $a=-\frac{1}{12}$.

\subsection{Monodromy action} 

The monodromy acts for the compact Calabi-Yau manifold $W$ as a subgroup of $SP(h_3(W),\mathbb{Z})$ 
on the CY periods $(F_I=\partial_I F,X^I)^T$, i.e. as
\begin{equation}
\left(\begin{array}{c} \tilde F_I\\ \tilde X^I\end{array}\right)=\left(\begin{array}{cc}  
A_I^{\phantom{I} J}& B_{IJ}\\ C^{IJ}& D^I_{\phantom{I} J}\end{array}\right)
\left(\begin{array}{c}  F_J\\ X^J\end{array}\right)
\label{globalmonodromy}
\end{equation}
with  all entries of $A_I^{\phantom{I} J}, B_{IJ}, C^{IJ}$ and $D^I_{\phantom{I} J}$ integers and
\begin{equation}
\left(\begin{array}{ll}  A^S_{\phantom {S}K}& (C^{T})^{SK}\\[2 mm]  (B^T)_{SK}& D_T^{\phantom{S}K}\end{array}\right)
\left(\begin{array}{cc}  0& -\delta^K_{\phantom{K} I}\\[2 mm] \delta_K^{\phantom{K} I}&0\end{array}\right)
\left(\begin{array}{ll} A_I^{\phantom{I} P}& B_{IP}\\[2 mm] C^{IP}& D^I_{\phantom{I} P}\end{array}\right)=
\left(\begin{array}{cc}  0& -\delta^S_{\phantom{S} P}\\[2 mm] \delta_S^{\phantom{S} P}&0\end{array}\right)\ .
\label{intersection}
\end{equation}
One clear advantage of the big phase space is that the monodromy acts simply on the tensors 
in the homogeneous coordinates. E.g. $\tau$ transforms as 
\begin{equation}
\tilde \tau_{IJ}=(A\tau +B)_{IK}(C\tau+D)^{-1\ K}_{\phantom{-1\ K} J}\
\label{tautrans1} 
\end{equation}  
and modular objects of tensor weight $-N$ transform like 
$\tilde C_{I_1,\ldots,I_n} = (C\tau+D)^{-1\ K_1}_{I_1},\ldots, (C\tau+D)^{-1\ K_N}_{I_N} C_{K_1,\ldots,K_N}$.

The monodromy for the non-compact cases acts on the periods 
$\Pi^T=({\cal F}_i=\int_{b_i}\lambda ,t^i=\int_{a^i} \lambda,m_\mu=\int_{\gamma_\mu} \lambda)$,
where $a^i,b_i$ is a symplectic basis of $H_1({\cal C}_g,\mathbb{Z})$ and $\gamma_\mu$ are
cycles encircling the points where $\lambda$ has a pole with non-vanishing residua.
As mentioned above $\Pi^T$ can be obtained as the periods of $W$ which stay finite 
in the non-compact limit\footnote{Typically the fundamental period $X^0$ in the large radius 
limit becomes one of the constant periods $m_\mu$ see e.g. the discussion of 
local ${\cal O}(-3)\rightarrow \mathbb{P}^2$ in the large fiber limit of  
the elliptic fibration over $\mathbb{P}^2$ realized as  $X_{18}(1,1,1,6,9)$, 
see \cite{Haghighat:2008gw}.}. We call ${\cal C}_g^*={\cal C}_g\setminus \{ p_i\}$. 
The monodromy acting on $H_1({\cal C}_g,\mathbb{Z})$ is a subgroup of $SP(2g,\mathbb{Z})$.
The action on $\Pi$ is
\begin{equation}
\left(\begin{array}{c} \tilde {\cal F}_i\\ \tilde t^i\\ m_\mu\end{array}\right)=\left(\begin{array}{ccc}  
a_i^{\phantom{i} j}& b_{ij}& l_{i\mu}\\ c^{ij}& d^i_{\phantom{i} j}&  l^i_\mu \\
0&0&\mathbb{I} \end{array}\right)
\left(\begin{array}{c}  {\cal F}_j\\ t^j \\ m_mu \end{array}\right)
\label{localmonodromy}
\end{equation}
and analogous to (\ref{intersection}) we have from the preservation of the intersection form
$a^Tc=c^ta$, $b^T d=d^Tb$ and $a^T d-c^Tb=\mathbb{I}$, with all entries of $a,b,c,d$ and 
$l_{i\mu},l^i_\mu$ integer. If ${\cal C}_g^*$ is obtained by a non-compact limit from 
$W$ the monodromy group of ${\cal C}_g^*$ generated by~\ref{localmonodromy} 
is a subgroup of the monodromy group of $W$. The action on $\tau_{ij}$ is
given similarly as in (\ref{tautrans1}) by 
\begin{equation} 
\tilde \tau_{ij}=(a\tau +b)_{ik}(c\tau+d)^{-1\ k}_{\phantom{-1\ k} j}\ .
\end{equation}    

An important difference is in the properties of the matrix $\tau$.
In the global case ${\rm Im}(\tau_{IJ})$ has signature $(1,h_{21})$, i.e. one negative 
eigenvalue. On the other hand as it was mentioned in section (\ref{sectionmasslessdirect}) it is a key
property of the solution of~\cite{SW1,SW2,KLT} that ${\rm Im}(\tau_{ij})$,
$i=1,\ldots,{\rm rank}(G)$, with $\tau_{ij}=\frac{i}{2 \pi} \partial_{a^i} \partial_{a^j} F^{(0)}$   
is positive definite. Mathematically $\tau_{ij}$ defines the Siegel upper half space associated to ${\cal C}_g$.  
In the non-compact limit the matrix ${\rm Im} (\tau_{IJ})$ is therefore projected a positive 
definite submatrix. 

The Hodge star operator $*$ on $W$ defines a natural complex structure on $H^3(W)$, 
which is $+i$ on $H^{3,0}\oplus H^{1,2}$ and $-1$ on  $H^{2,1}\oplus H^{0,3}$. 
This leads to the so called Weil intermediate Jacobian, which comes with a natural 
pairing given by the an-holomorphic matrix
\begin{equation} 
N_{IJ}=\tau_{IJ}-2 i \frac{{\rm Im}(\tau_{IK}){\rm Im}(\tau_{JL}) \bar X^L \bar X^K}{{\rm Im}( \tau_{KL})\bar X^L \bar X^K}\ .
\end{equation}
It is well know in supergravity that this defines the matrix of theta angles and the gauge 
couplings as $N_{IJ}=:\frac{\Theta_{IJ}}{\pi} +8 \pi i (g^{-2})_{IJ}$ and that 
${\rm Im}(N_{IJ})$ is positive. In the non-compact limit certain a submatrix of the anholomorphic  
$N_{IJ}$ becomes the holomorphic $\tau_{ij}$ of the rigid gauge theory. The Griffith complex 
structure on $H^3(W)$ is defined by $+i$ on $H^{3,0}\oplus H^{2,1}$ and $-1$ on $H^{1,2}\oplus H^{0,3}$ 
and the paring is given by $\tau_{IJ}$, which as mentioned above has one negative eigenvalue. 
We note that  $\tau_{IJ}$ and $N_{IJ}$ transform in the same way under $SP(h_3(W),\mathbb{R})$ 
transformations.

\section{Matrix model approach}
\label{matrixsection-05-19}

The study of the gravitational couplings of $SU(2)$ Seiberg-Witten
theory has been a fruitful setting to explore various approaches
for solving topological expansions. As reviewed in the introduction 
one can obtain the gravitational coupling $F^{(g)}$ by 
geometric engineering from toric Calabi-Yau 3-folds and by
Nekrasov's instanton counting calculations. Both approaches, the former 
via the vertex formalism, lead to sums over partitions, which 
are valid  in one region in the moduli space. In the 
geometric engineering approach one has in addition to take a 
limit. The direct integration of holomorphic anomaly 
equation~\cite{Huang} studied in the previous sections 
yields an analytic description of the higher genus 
amplitudes, which is recurse in the genus, but valid 
throughout the moduli space. 

In this section we will turn to another approach, namely the 
matrix model method. The matrix model is in principle a
framework that encodes exact perturbative information and possible 
non-perturbative completions. It was pioneered by Dijkgraaf and Vafa in 
particular in~\cite{DV2}. Following these suggestions the 
authors of~\cite{KMT} computed the gravitational couplings of 
$\mathcal{N}=2$ Seiberg-Witten theory by a limit from the Hermitian 
matrix model describing the glueball superpotential of 
$\mathcal{N}=1$ gauge theory. However, only the genus one 
amplitude $F^{(1)}$ has been obtained in this way, and it is not clear 
how to compute higher genus amplitudes in this approach,  
because $F^{(g)}$ is not gauge invariant for $g>1$~\cite{KMT}.

A microscopic matrix model was recently derived in~\cite{Klemm:2008yu} 
from the partition function~\cite{Nekrasov1,Nekrasov2} using the matrix 
model descriptions of infinite partitions~\cite{Eynard:2008mt}. 
Motivated by the recent works of \cite{EO, EMO}, we will apply the
formalism in \cite{EO} to the topological expansion of $SU(2)$
Seiberg-Witten theory. The formalism of~\cite{EO} has been developed 
from the study of loop equations in matrix models. It also proceeds
recursively genus by genus. One advantage of the formalism is that
one no longer need to refer to a matrix model in this set up. The defining date 
are the spectral curve ${\cal C}$ and the differential $\lambda$, which 
yields the filling fraction and the open one point function. The 
Seiberg-Witten curve has been shown to be the spectral curve of the 
microscopic matrix model considered in~\cite{Klemm:2008yu}. It also 
follows from a double scaling limit of the spectral curve of 
Gross-Witten matrix model considered by Dijkgraaf and Vafa in~\cite{DV2}. 

One obvious advantage of the matrix model approach is that it 
gives also the open amplitudes. Given the local mirror curve ${\cal C}^*$ 
and the  meromorphic differential  $\lambda$ for topological string 
theory on local Calabi-Yau manifolds the matrix model 
predictions for the open amplitudes have been checked against 
the topological vertex results~\cite{Marino:2006hs,Bouchard:2007ys}. 
The interpretation of these amplitudes in the gauge theory context 
is less clear. 

\subsection{Review of the formalism}
Here we review the formalism developed by Eynard and Orantin for integrating 
the loop equation. For more details and references see~\cite{EO}.

The algorithm is particularly elegant for elliptic curves in 
Weierstrass form. We will therefore focus on Seiberg-Witten 
curves in the Weierstrass form
\begin{equation} \label{weierstrass}
y^2=4x^3-g_2(u)x-g_3(u)\ .
\end{equation}
Here is $u$ is the Coulomb modulus of the Seiberg-Witten theory.
For the massless $N_f=2$ theory we find by transforming (\ref{curves})
into Weierstrassform  
\begin{equation} \label{gnf=2}
g_2(u)=\frac{4}{3}(u^2+3),~~~~g_3(u)=\frac{8u}{27}(u^2-9)\ ,
\end{equation}
where the three roots of (\ref{weierstrass}) are $x=\frac{2u}{3}, 1-\frac{u}{3},
-1-\frac{u}{3}$ respectively. We note that is also the Weierstrass form for
the Seiberg-Witten curve for  pure $SU(2)$ gauge theory as quoted
in~\cite{SW1}, while if we transform the $N_f=0$ case~\cite{Klemm:1994qs} 
in (\ref{curves}) into Weierstrass form we obtain 
\begin{equation} \label{gnf=0}
g_2(\tilde u)=-1+\frac{4\tilde u^2}{3}, ~~~~~~g_3(u)=\frac{1}{27}(9\tilde
u-8\tilde u^3)\ .
\end{equation}
The two curves specified by (\ref{weierstrass}) with (\ref{gnf=2}) or
(\ref{gnf=0}) respectively are known to be isogeneous. That means in 
particular that the Picard-Fuchs equation are the same, but a careful
analysis of the integral basis of $H_1({\cal C}_1,\mathbb{Z})$ 
reveals the $b$ periods differ by a factor of two. The relation 
between $\tilde u$ and $u$ from the comparison of the $J$-function (\ref{J})
is $u=\frac{\tilde u}{\sqrt{\tilde u^2-1}}$, i.e. it exchanges the 
asymptotic free region and the monopole region~\cite{KLT}. At genus 
zero it is difficult to distinguish the curves. Since the  Picard-Fuchs 
equations are the same the holomorphic prepotential can be derived 
from any of them. However at genus  one there is an important difference.   
We know that for $N_f=0$ the conifold factor is $\Delta\sim (u^2-1)$, while for
the massless $N_f=2$ case it is  $\Delta\sim (u^2-1)^2$, see (\ref{F1nf2}) and 
(\ref{conifolddivisors}). By calculating $\Delta$ from (\ref{Delta}), we see
that (\ref{weierstrass},\ref{gnf=2}) is the $N_F=2$ curve. Now an important 
simplification for the application of the~\cite{EO} formalism arises if the 
meromorphic differential $\lambda$ is simple rational function of the 
Weierstrass ${\cal   P}$-function for the curve written in the Weierstrass 
form. It turns out that for the Weierstrass curve of $N_f=2$ the 
form of the differential $\lambda=\frac{\sqrt{2}}{2 \pi} y \frac{d x}{x^2-1}$ 
used in~\cite{SW1} for the cubic curve quoted there as $N_f=0$ curve  has this 
property, see (\ref{5.109-05-19}). On the other hand if we transform the 
meromorphic (\ref{lambda}) for $N_f=0$ to the Weierstrass 
representation we cannot express it as a rational function of 
the  Weierstrass  ${\cal P}$-function. As we mention above for 
genus zero prepotential it is not relevant to match precisely 
the correct pair of curve and differential, but for higher genus it is 
crucial. Below we stick to the technically simplest case  namely  
the $N_f=2$ case.

Given a curve ${\cal C}_g$ the associated Bergmann kernel is 
defined as the unique bilinear meromorphic form with a single pole
of degree 2, whose integral over the $A$-cycles vanish, see~\cite{EO} 
for details. For the family of genus one curves (\ref{weierstrass}), the
associated Bergmann kernel $B(p,q)$ is simply the Weierstrass
${\cal P}$-function plus a constant $X$ 
\begin{equation} \label{Bergmann-05-18}
B(p,q)=(\wp(p-q)+X)dpdq \ .
\end{equation}
The Weierstrass ${\cal P}$ function is a double periodic, even 
function on $\mathbb{C}$ 
\begin{equation}
\wp(p+2a_1)=\wp(p), ~~~~\wp(p+2a_2)=\wp(p),~~~~\wp(-p)=\wp(p)\ ,
\end{equation}
which has a double pole around the  origin and the series expansion
\begin{eqnarray}
\wp(p)=\frac{1}{p^2}+\frac{g_2}{20}p^2+\frac{g_3}{28}p^4+\mathcal{O}(p^6)\ .
\end{eqnarray}
In particular  the ${\cal P}$-function is well defined on the two torus 
${\cal C}_1=\mathbb{C}/\Lambda$, where $\Lambda$ is the lattice spanned 
by the periods $(2 a_1, 2 a_2)$. The complex structure of ${\cal C}_1$ 
is $\tau=\frac{a_2}{a_1}$ and $(a_1,a_2)$ are half periods.

The constant $X$ can be fixed by the A-cycle integral
\begin{equation}
\int_{0}^{2a_1}B(p,q)= (-2\zeta(a_1)+2a_1X)dp\ ,
\label{acycle}
\end{equation}
where $\zeta(p)$ is the Weierstrass zeta function, and its value at
half period is related to the second Eisenstein series of $\tau$ as
\begin{eqnarray}
\zeta(a_1)a_1=\frac{\pi^2}{12}E_2(\tau)\ .
\end{eqnarray}
Using the relations between Weierstrass invariants and the Eisenstein series
\begin{eqnarray} \label{5.100-05-18}
(2a_1)^4g_2(u)=\frac{4\pi^4}{3}E_4(\tau),~~~~(2a_1)^6g_3(u)=\frac{8\pi^6}{27}E_6(\tau)
\end{eqnarray}
and the vanishing of (\ref{acycle}), we determine $X$ in (\ref{Bergmann-05-18})
\begin{eqnarray} \label{cons-05-19}
X=\frac{3g_3(u)E_2(\tau)E_4(\tau)}{2g_2(u)E_6(\tau)}\ .
\end{eqnarray}
Because of (\ref{E2trans}) the Bergmann Kernel transforms with 
a shift under modular transformations. One can define 
the modular invariant modified Bergmann kernel by replacing  
$E_2$ in (\ref{Bergmann-05-18}) with $\hat{E}_2$, as 
defined in (\ref{E2hat}). This replacement induces an isomorphism 
between ring of quasimodular forms and the ring of almost 
holomorphic modular forms. In the manipulations below 
we can work with $E_2$ and replace it at the end of 
calculations with $\hat E_2$, if we wish to consider 
truly modular objects.

The Eisenstein series are related to Jacobi theta functions by the
well-known formulae
\begin{eqnarray}
E_4(\tau)&=&b^2+bd+d^2 \nonumber \\
E_6(\tau)&=&\frac{1}{2}(2d^3+3bd^2-3b^2d-2b^3) \ , 
\label{5.95-05-18}
\end{eqnarray}
where $b,c,d$ are defined in (\ref{defabcd}).  The modulus $u$
and half period $a_1$ can be written in terms of Jacobi theta
function using (\ref{5.95-05-18}) and (\ref{5.100-05-18}) 
as~\footnote{One can solve for
$u$ by eliminating $a_1$ in (\ref{5.100-05-18}). There are other
solutions besides the solution $u=1+\frac{2d}{b}$ we use. They
correspond to various special points in the Coulomb moduli space
as $\tau\rightarrow i\infty$, or are related the one we use by
$Z_2$ symmetry. Without loss of generality we will just use the
solution $u=1+\frac{2d}{b}$ in order to compare with large $u$,
i.e. weak coupling limit.}
\begin{eqnarray} \label{ua-05-19}
u=1+\frac{2d}{b},~~~a_1^2=\frac{\pi^2 b}{8}
\end{eqnarray}
Therefore the constant in (\ref{cons-05-19}) can be written in
terms of modular forms
\begin{eqnarray} \label{variableX-06-25}
X=\frac{2E_2(\tau)}{3b}\ .
\end{eqnarray}
In the Weierstrass form (\ref{weierstrass}) the Seiberg-Witten curve is 
parameterized by the Weierstrass function and its derivative via the identification
\begin{eqnarray}
y=\wp^{\prime}(p),~~~x=\wp(p) \ .
\end{eqnarray}
The branching points of the algebraic curve (\ref{weierstrass})
are the points of $dx=0$, which are simply the half periods $a_1,
a_2, a_3=a_1+a_2$ in the case of Weierstrass function. The values
of Weierstrass function at half periods are the roots of
Weierstrass equation $4x^3-g_2x-g_3=0$. The ordering will not be
important for us, so without loss of generality we can take
\begin{eqnarray}
\wp(a_1)=\frac{2u}{3},~~\wp(a_2)=1-\frac{u}{3},~~
\wp(a_3)=-1-\frac{u}{3}\ .
\end{eqnarray}
The derivative of Weierstrass function vanishes at the half
periods $\wp^{\prime}(a_1)=\wp^{\prime}(a_2)=\wp^{\prime}(a_3)=0$.
For a point $p$ near each branching point $a_i$, there is an
unique image denoted as $\bar{p}$ such that $\wp(p)=\wp(\bar{p})$.
Since the Weierstrass function satisfies $\wp(2a_i-p)=\wp(p)$, we
can easily determine
\begin{eqnarray}
\bar{p}=2a_i-p
\end{eqnarray}
Higher derivatives of Weierstrass function can be related to
Weierstrass function and its derivative algebraically, for example
we have the formula for the second derivative as
$\wp^{\prime\prime}(p)=-\frac{g_2}{2}+6\wp(p)^2$, etc.

The periods of Seiberg-Witten theory should correspond to the
``filling fraction'' defined in \cite{EO}. In the $N_f=2$  massless theory 
it is the integral of the following meromorphic differential 
\begin{eqnarray} \label{5.109-05-19}
\lambda(p)&=&\frac{1}{2\sqrt{2}}\frac{y(p)dx(p)}{(x(p)-\wp(a_2))(x(p)-\wp(a_3))}
\nonumber \\ &=&
\frac{1}{2\sqrt{2}}\frac{\wp^{\prime}(p)^2dp}{(\wp(p)-\wp(a_2))(\wp(p)-\wp(a_3))}
\end{eqnarray}
over the cycles
of algebraic curve, i.e. $a=\frac{1}{2\pi i}\int_a \lambda(p)$. 
Here we have chosen a normalization for which the derivative of the 
prepotential is $\frac{\partial^2 F^{(0)}}{\partial a^2}=-2\pi i \tau$. This 
will be convenient later on.

A set of diagrammatic rules are provided in \cite{EO} to construct
the topological expansion $F^{(g)}$ associated with the algebraic
curve. Below we list the basic components and their expansions around the
branching points $a_i, i=1,2,3$:
\begin{enumerate}
\item The vertex $\omega(p)$. This can be constructed from the differential
one-form in (\ref{5.109-05-19}) as the following,
\begin{eqnarray}
\omega(p) &=&
\frac{1}{2\sqrt{2}}\frac{(y(p)-y(\bar{p}))dx(p)}{(x(p)-\wp(a_2))(x(p)-\wp(a_3))}
\nonumber \\ &=&
\frac{1}{\sqrt{2}}\frac{\wp^{\prime}(p)^2dp}{(\wp(p)-\wp(a_2))(\wp(p)-\wp(a_3))}
\end{eqnarray}
It is straightforward to compute the series expansion near the
branching points. For the branching point $a_1$, the vertex
$\omega(p)$ goes like $\omega(p)\sim (p-a_1)^2$, while the other
two points $a_2$ and $a_3$, the vertex goes like $\omega(p)\sim
\mathcal{O}(1)$.

\item The root $\Phi(p)$. This is simply the integral of the differential
one-form $\lambda(p)$ in (\ref{5.109-05-19}) from any base point
on the algebraic curve
\begin{eqnarray}
\Phi(p) &=& \int^p \lambda(p) \ .
\end{eqnarray}
The integration constant will not appear
in final answers and will not be important.

It is straightforward to compute the series expansion of
$\lambda(p)$ near the branching points and perform the integral.
At the branching point $a_1$ the root $\Phi(p)$ has the leading 
behaviour $\Phi(p)\sim (p-a_1)^3$, while at the other two points 
$a_2$ and $a_3$, it behaves like $\Phi(p)\sim (p-a_i)$.

\item The line-propagator is simply the Bergmann kernel
$B(q,p)$. We expand it in the first variable $q$ around a
branching point $a_i$,
\begin{eqnarray}
\frac{B(q,p)}{dpdq} &=& \wp(a_i-p)+X+\wp^{\prime}(a_i-p)(q-a_i) \nonumber \\
&&+ \frac{1}{2}\wp^{\prime\prime}(a_i-p)(q-a_i)^2
+\mathcal{O}((q-a_i)^3)\ .
\end{eqnarray}
We then expand in the second variable $p$ around another
branching point $a_j$. If $a_i= a_j$, there will be poles as
$p\rightarrow a_i$. For $a_i\neq a_j$, there will be no pole. 
In both cases it is straightforward obtain the series expansions.

\item The arrow-propagator $dE_q(p)$ is an integral of the
Bergmann kernel and can be expanded around a branching point $a_i$
in the following way
\begin{eqnarray}
\frac{dE_q(p)}{dp} &=& \frac{1}{2}\int_q^{\bar{q}}B(\xi,p) \nonumber \\
&=&
-(\wp(a_i-p)+X)(q-a_i)-\frac{1}{6}\wp^{\prime\prime}(a_i-p)(q-a_i)^3
\nonumber \\ && -\frac{1}{120}\wp^{(4)}(a_i-p)(q-a_i)^5 +\cdots
\end{eqnarray}
Again, if $a_i= a_j$, there will be poles  as $p\rightarrow
a_i$, otherwise for $a_i\neq a_j$, there will be no pole. The necessary
series expansions are straightforward to obtain.
\end{enumerate}
\medskip

{}From these basic components one can construct the correlation
functions $W^{(g)}_k(p_1,\cdots,p_k)$, and free energy $F^{(g)}$
for all $g\geq2$ in terms of some residue formulae. For example,
The genus one  one-point function is
\begin{eqnarray} \label{5.144-05-19}
W^{(1)}_1(p)= \mathop{\rm Res}_{q\rightarrow
\bf{a}}\frac{dE_q(p)}{\omega(q)}B(q,\bar{q})
\end{eqnarray}
and the genus two free energy $F^{(2)}$ is
\begin{eqnarray} \label{5.145-05-19}
F^{(2)}&=& -\frac{1}{2}\mathop{\rm Res}_{p\rightarrow
\bf{a}}\mathop{\rm Res}_{q\rightarrow \bf{a}}\mathop{\rm
Res}_{r\rightarrow \bf{a}}\mathop{\rm Res}_{s\rightarrow
\bf{a}}\{\frac{\Phi(p)dE_q(p)}{\omega(q)}\frac{dE_r(q)}{\omega(r)}\frac{dE_s(\bar{q})}{\omega(s)}
B(r,\bar{r})B(s,\bar{s}) \nonumber \\ && +
\frac{\Phi(p)dE_q(p)}{\omega(q)}\frac{dE_r(q)}{\omega(r)}\frac{dE_s(\bar{r})}{\omega(s)}
B(r,\bar{q})B(s,\bar{s})+
\frac{\Phi(p)dE_q(p)}{\omega(q)}\frac{dE_r(q)}{\omega(r)}\frac{dE_s(r)}{\omega(s)}
\nonumber \\ && \times
[B(\bar{r},\bar{q})B(s,\bar{s})+B(\bar{s},\bar{q})B(s,\bar{r})+B(s,\bar{q})B(\bar{s},\bar{r})]
\}\ ,
\end{eqnarray}
where the residues are taken around the three branching points
$a_1, a_2, a_3$.

\subsection{Calculations of open and close amplitudes}

We calculate the genus one  one-point function $W^{(1)}_1(p)$ and
the genus two free energy $F^{(2)}$ for the Seiberg-Witten curve
(\ref{weierstrass}). As we mentioned it describes the $SU(2)$
Seiberg-Witten theory with two massless flavors. The genus one  one-point function $W^{(1)}_1(p)$ is calculated
from (\ref{5.144-05-19}), we find
\begin{eqnarray}
W^{(1)}_1(p)&=& \sum_{i=1}^3\mathop{\rm Res}_{q\rightarrow
a_i}\frac{dE_q(p)}{\omega(q)}B(q,\bar{q}) \nonumber \\
&=&
-\frac{dp}{48\sqrt{2}(u^2-1)}[4(u-6X)(\wp(p-a_1)+X)+6(u-1)(\wp(p-a_2)+X)
\nonumber
\\ &&+6(u+1)(\wp(p-a_3)+X)-\wp^{\prime\prime}(p-a_1)]\ . \label{W11-06-26}
\end{eqnarray}
The genus one free energy $F^{(1)}$ is not directly constructed
from the diagrammatic rules, but the derivative of it with respect
to the Seiberg-Witten period is the integral of $W^{(1)}_1(p)$
over the B-cycle
\begin{eqnarray}
\frac{\partial F^{(1)}}{\partial a}=\int_{0}^{2a_2}W^{(1)}_1(p)\ .
\end{eqnarray}
Using the formulae for Weierstrass zeta function
$2a_2\zeta(a_1)-2a_1\zeta(a_2)=\pi i$ and (\ref{ua-05-19}) we can
compute the integral
\begin{eqnarray} \label{F1a-06-26}
\frac{\partial F^{(1)}}{\partial
a}=\int_{0}^{2a_2}W^{(1)}_1(p)=\frac{i\sqrt{b}}{6cd}(E_2-b-2d)\ .
\end{eqnarray}
This matches with our earlier calculations for Seiberg-Witten
theory with $N_f=2$ massless flavors, using (\ref{a-05-19}),
(\ref{F1-05-19}) \footnote{There is an extra factor of $i$
comparing with (\ref{a-05-19}), (\ref{F1-05-19}). This is because
the matrix model should describe the expansion of $F^{(g)}$ around
the conifold point for which the filling fraction is real and goes
to zero, instead of the point $u\rightarrow \infty$. The formulae
(\ref{ua-05-19}) we have used are for the point at infinity
$u\rightarrow \infty$, and should become $u=1+\frac{2b}{d}$, $
a_1^2=-\frac{\pi^2 d}{8}$ for the conifold point. The extra factor
of $i$ is then cancelled due to the extra minus sign of $a_1^2$.
Since this problem will not appear at higher genus $g\geq 2$, we
will still use the convention at $u\rightarrow \infty$ for
convenience in comparing with instanton counting. }.

Now we come to genus two free energy, we compute the various terms
in (\ref{5.145-05-19}) and the total result is
\begin{eqnarray}
F^{(2)}=\frac{675X^3-1350uX^2+(990u^2+1350)X-16u^3-1080u}{6480(u^2-1)^2}\ .
\end{eqnarray}
Substituting in $u=1+\frac{2d}{b}$ and $X=\frac{2E_2}{3b}$, we
find the agreement with earlier calculations (\ref{Nf2masslessF2})
for Seiberg-Witten theory with $N_f=2$ massless flavors using
holomorphic anomaly.

Similarly, we can compute the genus two one-point function 
\begin{eqnarray} \label{eq-6.142-06-25}
W^{(2)}_1(p) &=& \frac{5}{32\sqrt{2}(u^2-1)^3}\big{[}X^5+\frac{6u\wp(p)+3-7u^2}{2u-3\wp(p)}X^4 \nonumber \\ && 
+\frac{9(75+77u^2)\wp(p)^2-12u(75+77u^2)\wp(p)+2(405-660u^2+559u^4)}{45(2u-3\wp(p))^2}X^3
\nonumber \\  && + b_2X^2+b_1X+b_0
\big{]}
\end{eqnarray}
where $b_0,b_1,b_2$ are some very complicated functions of the Weierstrass
function $\wp(p)$ and $u$\footnote{They are too cumbersome to write down here,
  but are available upon request.}. 
Some empirical remarks can be made about a genus $g$ one-point amplitude $W^{(g)}_1(p)$:
\begin{enumerate}
\item $W^{(g)}_1(p)$ is a polynomial of $X$ of degree $3g-1$.
\item The coefficients of the polynomial are rational functions of  $\wp(p)$
  and $u$. They are regular at $p=0$ (or equivalently $\wp(p)=\infty$). They
  are singular at the half periods $p=a_1,a_2, a_3$. The degree of poles of $\wp(p)-\wp(a_1)$,
$\wp(p)-\wp(a_2)$, $\wp(p)-\wp(a_3)$ are $g+3$, $2$, $2$ respectively. 
For example, the coefficient $b_0$ in (\ref{eq-6.142-06-25}) as a rational function $\wp(p)$ can be written as 
\begin{eqnarray}
b_0 &=& \frac{A(\wp(p))}{(\wp(p)-\wp(a_1))^5(\wp(p)-\wp(a_2))^2(\wp(p)-\wp(a_3))^2}
\nonumber \\ &\sim& \frac{A(\wp(p))}{(3\wp(p)-2u)^5(9\wp(p)^2+6u\wp(p)+u^2-9)^2}
 \end{eqnarray} 
 where $A(\wp(p))$ is a polynomial of $\wp(p)$ of degree $9$.   
 \end{enumerate}
The boundary behavior of close string moduli $u$ especially at the conifold
point $u\rightarrow 1$ is discussed  more details in 
section~\ref{openboundarybehaviour}.

\subsection{Holomorphic anomaly equation for open amplitudes}
We see that we can use the matrix model formalism to compute  higher genus
topological amplitude for the massless $N_f=2$ Seiberg-Witten theory. But the 
formalism gets quite complicated at higher genus, and for the close
topological amplitude $F^{(g)}$, the most efficient way of calculation is
still through the use the holomorphic anomaly equation plus boundary
conditions at the conifold point. One might wonder whether this method of 
``direct integration'' can also be applied to the open topological amplitude. 
In order to explore this idea, we consider a version of the holomorphic
anomaly equation for the open topological amplitudes proposed in~\cite{EMO} 
based on the matrix model formalism. 

An extended open holomophic anomaly equation has been applied to the 
calculations of open amplitudes on the  the real quintic Calabi-Yau 
manifold~\cite{Walcher:2007tp}. This formalism was recently applied 
to local ${\cal O})(-3)\rightarrow \mathbb{P}^2$~\cite{Krefl:2009md}.
It differs from the discussion here, as it encorporates no open moduli. 

The open holomorphic anomaly equation of \cite{EMO} is 
\begin{eqnarray} \label{EMO-06-25}
\partial_{\bar{K}}W^{(g)}_k=
\frac{1}{2}C^{IJ}_{\bar{K}}(D_ID_JW^{(g-1)}_k+\sum_h\sum_{L\subset K} D_I
W^{(h)}_lD_J W^{(g-h)}_{k-l})\ ,
\end{eqnarray}
where the $I, J, K$ are close string moduli. For our toy model of $N_f=2$
$SU(2)$ Seiberg-Witten theory, the only anti-holomorphic dependence comes from 
the function $\hat{E}_2(\tau)$, which appears in the variable $X$ we defined  
in (\ref{variableX-06-25}). The close string moduli in this case can be
parametrized by the period $a$, and since it is a flat coordinate in the 
holomorphic limit, the covariant derivatives in the RHS of (\ref{EMO-06-25}) 
can be replaced by just ordinary derivatives. After fixing the normalization
correctly, the equation (\ref{EMO-06-25}) becomes for the case at hand  
\begin{eqnarray}
-\frac{16}{\theta_2^4(\tau)}\frac{\partial W^{(g)}_k}{\partial X}=
\partial^2_a W^{(g-1)}_k
+\sum_h\sum_{L\subset K} \partial_a W^{(h)}_l\partial_a W^{(g-h)}_{k-l} \ .
\end{eqnarray}

Consider the simplest case of the above open holomorphic anomaly equation, 
namely the case $g=0$ and $k=3$. The equation becomes
\begin{eqnarray}
-\frac{16}{\theta_2^4(\tau)}\frac{\partial W^{(0)}_3(p,q,r)}{\partial X} &=&
2\partial_a W^{(0)}_1(p)\partial_a W^{(0)}_2(q,r) +2\partial_a
W^{(0)}_1(q)\partial_a W^{(0)}_2(p,r)\nonumber \\ &&  +2\partial_a
W^{(0)}_1(r)\partial_a W^{(0)}_2(p,q) \ . \label{eq6.146-06-27}
\end{eqnarray}
To test the equation, we can use the residue formulae to  compute directly 
the genus zero 3-point function 
\begin{eqnarray} \label{W03-06-27}
W^{(0)}_3(p,q,r)& = & \mathop {\rm Res}_{s\rightarrow {\bf a}} \frac{dE_s(p)}{\omega(s)}[B(s,q)B(\bar{s},r)+B(\bar{s},q)B(s,r)]
\nonumber \\ & =&
-\frac{(\wp(p-a_1)+X)(\wp(q-a_1)+X)(\wp(r-a_1)+X)}{\sqrt{2}(u^2-1)}\ . 
\end{eqnarray}
On the RHS, the genus zero one-point function is undefined in the matrix model
formalism, and the notation of $\partial_a W^{(0)}_1(p)$ simply means the
contour integral of $W^{(0)}_2(p,q)=B(p,q)$ over the B-cycle. We find
\begin{eqnarray}
\partial_a W^{(0)}_1(p)=\int^{2a_2}_0 B(p,q) dq=\frac{\pi i}{a_1}=\frac{2\sqrt{2}i}{\theta^2_2(\tau)}
\end{eqnarray}
We see an immediate problem with (\ref{eq6.146-06-27}). The LHS has a pole at
$p\rightarrow a_1$, but the RHS involves the Weierstrass Zeta function from 
$\partial_a \wp(p-q)$ and does not have a pole at $p\rightarrow a_1$. The 
discrepancy  comes from the fact that in the derivation of the open
holomorphic anomaly equation \cite{EMO}, the contour integral is converted
into covariant derivative of the close string moduli. However, it seems that
this procedure is not valid in the presence of open string moduli, so we have
to do the contour integral directly instead of just taking derivative. Namely, 
\begin{eqnarray} \label{eq6.154-11-24}
\partial_aB(p,q)\neq \int_0^{2a_2} W^{(0)}_3(q,r,s) ds
\end{eqnarray}
So the {\it correct} version of the open holomorphic anomaly equation (\ref{eq6.146-06-27}) should be 
\begin{eqnarray} \label{eq6.149-06-27}
-\frac{16}{\theta_2^4(\tau)}\frac{\partial W^{(0)}_3(p,q,r)}{\partial X}=2
\partial _a W^{(0)}_1(p)  \int_0^{2a_2} W^{(0)}_3(q,r,s)ds 
+ \textrm{permutation}.  
\end{eqnarray}
We check this is indeed satisfied by plugging in the expression for genus zero
3-point function (\ref{W03-06-27}). However, this is not much useful for the 
purpose of computing $W^{(0)}_3(p,q,r)$ as it appears in both RHS and LHS.

We also consider the case $g=1$ and $k=1$. Using (\ref{W11-06-26}). We get
\begin{eqnarray}
-\frac{16}{\theta_2^4(\tau)}\frac{\partial W^{(1)}_1(p)}{\partial X}= \frac{4\sqrt{2}}{3\theta^4_2(\tau)(u^2-1)}
[-3X+2u-3(\wp(p-a_1)+X)]
\end{eqnarray}
Again the naive equation 
\begin{eqnarray}
-\frac{16}{\theta_2^4(\tau)}\frac{\partial W^{(1)}_1(p)}{\partial X}=
\partial^2_a W^{(0)}_1(p)+2 \partial_a W^{(0)}_1(p)\partial_a F^{(1)} 
 \label{eq6.146-06-26}
\end{eqnarray}
is not correct,  as it can be seen that the RHS is independent of the open string modulus $p$ while the LHS is dependent on $p$.  The correct equation is 
\begin{eqnarray} \label{eq6.149-06-26}
-\frac{16}{\theta_2^4(\tau)}\frac{\partial W^{(1)}_1(p)}{\partial X}=- \int_0^{2a_2} \int_0^{2a_2}
W^{(0)}_3(p,q,r) dqdr+2    \partial _a W^{(0)}_1(p)  \partial_a F^{(1)} 
\end{eqnarray}
where the minus sign in the first term of RHS is just due to the different
conventions of using modular forms around conifold or infinity, and in the
second term the derivative  $\partial_a F^{(1)} $  is equal to the contour
integral of genus one-point function since there is no open string moduli. 

We summarize the findings in a few remarks.
\begin{enumerate}
\item The holomorphic anomaly equation (\ref{EMO-06-25}) is oversimplified and
the improved version does not seem to be too useful in computing higher point 
function $W^{(g)}_k$ when $k\geq 2$, because $W^{(g)}_k$ appears in 
both sides of the equation as exemplified by
(\ref{eq6.149-06-27}).   
\item The reason for this subtlety in (\ref{eq6.154-11-24}) is because we are
  using a non-standard differential one-form (\ref{5.109-05-19}) necessary for
  our calculations in $SU(2)$ Seiberg-Witten theory. If we used the standard
  differential one-form $\lambda=ydx$ as the \cite{EO}, the open holomorphic
  anomaly equation would be valid, but this would not be the right
  differential one-form to compute the gravitational coupling of Seiberg-Witten theory.
\item For the free energy $F^{(g)}$ and one-point function $W^{(g)}_1(p)$, the
  holomorphic anomaly equation can be used to determine the amplitudes up to a
  holomorphic anomaly. Only lower genus open amplitudes appear in the RHS of
  the holomorphic anomaly equation.  For example, in order to compute the
  genus two one-point amplitude $W^{(2)}_1(p)$ this way, we first have to 
 determine lower amplitudes up to $F^2$, $W^{(1)}_3(p_1,p_2,p_3)$.
\end{enumerate}

\subsection{Boundary condition for open topological amplitudes}
\label{openboundarybehaviour}
We now turn to another important issue of boundary conditions.  We consider
the limiting behavior open topological amplitudes around the conifold point, 
which is the point where $u\rightarrow 1$, $\tau_D=-\frac{1}{\tau}\rightarrow i\infty$, and  
\begin{eqnarray}
a_D=-\frac{i}{3\theta^2_4(\tau_D)}(E_2(\tau_D)-\theta_3^4(\tau_D)-\theta_2^4(\tau_D))\rightarrow
0\ .
\end{eqnarray}

We now expand genus one one-point function (\ref{W11-06-26}) around the
conifold point in terms of the flat coordinate $a_D$. Firstly it is 
convenient to rewrite the expression in terms of only $\wp=\wp(p)$, $u$, and $X$
\begin{eqnarray} \label{eq6.152-06-26}
W^{(1)}_1(p) &=& \frac{1}{2\sqrt{2}(u^2-1)}\bigl\{X^2+
\frac{3-3u^2}{2u-3\wp}X+\frac{1}{36(2u-3\wp)^2(9\wp^2+6u\wp+u^2-9)}\times \nonumber \\
&& [405u^2-1701)\wp^4+(216u^3+648u)\wp^3+(594u^4+1620u^2-486)\wp^2 \nonumber \\ 
&& +(384u^5-4896u^3+2592u)\wp+65u^6+501u^4+675u^2-729]\bigr\} \ .
\end{eqnarray}
We notice the Weierstrass function $\wp(p)$ is also dependent on the
underlying elliptic curve. However, for generic value of the open string
modulus $p$, the function $\wp(p)$ has a finite generic value at the conifold 
point of the close string moduli space. So we can first expand $u$ and $X$ in 
the expression  (\ref{eq6.152-06-26}), and treat $\wp(p)$ as an independent
parameter. Naively, we should expect the singular behavior as 
\begin{eqnarray}
W^{(1)}_1(p)= \mathcal{O}(\frac{1}{a_D})\ .
\end{eqnarray}
{\it Surprisingly}, we find that the leading singular term vanishes, and the
conifold expansion is regular. The series expansion result is 
\begin{eqnarray}
W^{(1)}_1(p) = \frac{3\wp(p)-2}{8\sqrt{2}(3\wp(p)+4)}+
\frac{i(9\wp(p)^2+24\wp(p)-8)}{8\sqrt{2}(3\wp(p)+4)^2}a_D+\mathcal{O}(a_D^2)\ .
\end{eqnarray}
Thus the regularity of the conifold expansion in this case imposes boundary
conditions for the open holomorphic ambiguity. For the genus one one-point
function (\ref{eq6.152-06-26}), the terms in the first line are fixed by the
open holomorphic anomaly equation (\ref{eq6.149-06-26}), and the rest is the
ambiguity which can be parametrized by 14 constants in this
case. Unfortunately, for generic holomorphic ambiguity, the coefficient of the
singular $\frac{1}{a_D}$ term  in the conifold expansion of $W^{(1)}_1(p)$
turns out to be a rational function of $\wp(p)$ whose numerator is a degree 4 polynomial of
 $\wp(p)$. So the conifold boundary condition only fixes 5 of the 14 unknown
constants in the holomorphic ambiguity of $W^{(1)}_1(p)$. More ingenuity may
be needed to completely fixes the holomorphic ambiguity. 

We also similarly test the conifold expansion of the genus two one-point
amplitude $W^{(2)}_1(p)$ in (\ref{eq-6.142-06-25}). The leading singular term
with generic holomorphic ambiguity is $\mathcal{O}(\frac{1}{a_D^3})$, but we
again find that the actual series is not singular
\begin{eqnarray}
W^{(2)}_1(p) =\frac{27(3\wp-2)}{512\sqrt{2}(3\wp+4)^2}+\frac{9i(27\wp^3
+216\wp^2+288\wp-224)}{1024\sqrt{2}(3\wp(p)+4)^3}a_D+\mathcal{O}(a_D^2)\ . \nonumber \\ 
\end{eqnarray}

\section{Future directions}

We have solved the topological sector of the $N=2$ $SU(2)$ gauge 
theories with $N_f=0,1,2,3$ matter multiplets in the fundamental
representation. Near the asymptotic free region in the vector multiplet 
space our results agree with the instanton calculation of Nekrasov. 
At the conifold points and the conformal points our globally 
defined expressions predict the topological sector of these 
theories in canonical holomorphic coordinates. 

Especially the analysis at the conformal points relies on the method 
proposed in~\cite{Huang}. It would be challenging and interesting 
to find a microscopic description especially at these points, at which the theory 
does not allow for an action formulation. The structure of the 
$F^{(g)}$ is very similar as at orbifold singularities in 
topological string theory~\cite{HKQ,Haghighat:2008gw,Alim:2008kp}, which 
suggests that a dual string description is a serious candidate.

We described the construction of the modular objects 
entirely from the Picard-Fuchs system in a form 
that generalizes straightforwardly to $N=2$ theories with
higher rank gauge groups and does not require knowledge of modular 
forms w.r.t. subgroups of $SP(2g,\mathbb{Z})$. E.g. the solutions 
for the periods of~\cite{KLT} for $SU(3)$ could be used to study the 
topological theory at Argyres-Douglas conformal points in $SU(3)$ theory.           
     
We find additional evidence that the simple boundary 
conditions namely the gap at the conifold and regularity of 
the amplitudes at the conformal points fix the entire ambiguity 
of $N=2$ topological theories associated to Riemann surfaces.
However one should prove integrability of these type of 
topological theories in general. 

Note that for the massless $N_f=4$ case the $F^{(g)}$ can 
be written as quasi-modular forms of weight $2g-2$ of 
${\rm PSL}(2,\mathbb{Z})$~\cite{Grimm:2007tm} similar 
as the $F^{(g)}$ for the asymptotic free cases here, 
but there is no gap structure in the conformal cases. 
It seems possible but tedious to fix the ambiguity 
here by considering mass perturbations and the associate 
limits to the cases that are treated in this paper.   

In the global case the above mentioned boundary conditions 
are not sufficient. We hope that this can be overcome by 
the study of various limit in multi-moduli compact 
Calabi-Yau manifold. For this reason we described 
the limit of rigid special K\"ahler geometry in great 
detail. Enough field theory limits, which are 
integrable, could make the global theory eventually 
also solvable.      

We also compared our calculation with the matrix 
model, respectively spectral curve approach of
Eynard and Orantin. This yields an alternative way to solve these 
theories, which  gives additional information about certain 
open matrix model amplitudes, whose meaning has not been 
studied in the context of gauge theory yet. 

In~\cite{Klemm:2008yu}  a microscopic matrix model for the 
Seiberg-Witten theory was derived starting from the instanton sums in asymptotic 
free regions. Here we go the opposite way and derive from 
the improved recursive formalism of~\cite{EO} the global higher amplitudes, 
whose expansion in the asymptotic region checks with~\cite{Nekrasov1}. 
Given the by now well established relation of Seiberg-Witten gauge theory 
with the matrix model makes the gauge theory  a most interesting laboratory 
to test the physical implications of the non-perturbative ideas that were 
recently put forward in the matrix model context~\cite{Eynard:2008he,Marino:2008vx}.

\vspace{0.2in} {\leftline {\bf Acknowledgments:}}

We thank Thomas Grimm, Babak Haghighat, Marcos Marino, Nicolas Orantin 
and Marco Rauch for fruitful discussions. 

\appendix

\section{Nekrasov's calculations}
In \cite{Nekrasov1} Nekrasov compute the Seiberg-Witten
prepotential and its gravitational corrections by instanton
counting. The results are represented by partition of instanton
number into Young tableau. The results for $SU(2)$ theory with one
massive hypermultiplet, i.e. $N_f=1$, up to 5-instanton and genus
2 are \footnote{Our convention has a sign difference from that of
\cite{Nekrasov1, Nekrasov2} at odd genus.}
\begin{eqnarray}
F^{(0)}&=& 4a^2\textrm{log}(a)+(c^{(0)}_2a^2+c^{(0)}_1a+c^{(0)}_0)
\nonumber
\\&&
-\frac{1}{2}(a+m_1)^2\textrm{log}(a+m_1)-\frac{1}{2}(-a+m_1)^2\textrm{log}(-a+m_1)
\nonumber \\ &&
-\frac{m_1}{2a^2}+\frac{3a^2-5m_1^2}{64a^6}+\frac{7a^2m_1-9m_1^3}{192a^{10}}
-\frac{153a^4-1430a^2m_1^2+1469m_1^4}{32768a^{14}} \nonumber \\
&&-\frac{1131a^4m_1-5250a^2m_1^3+4471m_1^5}{81920a^{18}} \label{NekrasovF0}\\
F^{(1)}&=&\frac{1}{12}\textrm{log}(\frac{(2a)^2}{a^2-m_1^2})+c^{(1)}-
\frac{3a^2-4m_1^2}{128a^8}-\frac{27a^2m_1-32m_1^3}{384a^{12}}\nonumber \\
&& +\frac{9(73a^4-733a^2m_1^2+732m_1^4)}{32768a^{16}}
+\frac{1899a^4m_1-9259a^2m_1^3+7848m_1^5} {16384a^{20}}
\label{NekrasovF1} \\
F^{(2)}&=&
-\frac{1}{480a^2}+\frac{1}{240(a+m_1)^2}+\frac{1}{240(-a+m_1)^2}
\nonumber \\ &&
+\frac{9a^2-11m_1^2}{1024a^{10}}+\frac{103a^2m_1-117m_1^3}{1024a^{14}}
-\frac{3(5583a^4-58186a^2m_1^2+57067m_1^4)}{262144a^{18}}
\nonumber \\
&& -\frac{451719a^4m_1-2273690a^2m_1^3+1919923m_1^5}{655360a^{22}}
\label{NekrasovF2}
\end{eqnarray}
Here $m_1$ is the mass of the hypermultiplet and for convenience
we have set the $\mathcal{N}=2$ dynamical scale $\Lambda=1$, which
can be easily recovered by dimensional analysis. The constants
$c^{(g)}_i$ are not important for us. In the above formulae we
have also included the leading perturbative terms. In $SU(2)$ case
the leading perturbative term at genus $g$ is \cite{Nekrasov1,
Nekrasov2}

\begin{eqnarray}
F^{(g)}_{pert}=\gamma_g(2a)+\gamma_g(-2a)-\sum_{i=1}^{N_f}\gamma(a+m_{i})-\sum_{i=1}^{N_f}\gamma(-a+m_{i})
\end{eqnarray}
where
\begin{eqnarray}
\gamma_0(x)&=&\frac{1}{2}x^2\textrm{log}(x)-\frac{3}{4}x^2
\nonumber
\\
\gamma_1(x)&=&\frac{1}{12}\textrm{log}(x) \nonumber \\
\gamma_2(x)&=&-\frac{1}{240}\frac{1}{x^2} \nonumber \\
\cdots \nonumber \\
\gamma_g(x)&=&
\frac{(-1)^gB_{2g}}{2g(2g-2)}\frac{1}{x^{2g-2}},~~g>1 \nonumber
\end{eqnarray}

We will also consider the much simpler case where the
hypermultiplets are massless. We list the results for $N_f=1,2,3$
and up to 5-instanton, genus 3. For $N_f=1$
\begin{eqnarray} \label{NekrasovmasslessNf1}
F^{(0)}&=&
3a^2\textrm{log}(a)+(c^{(0)}_2a^2+c^{(0)}_1a+c^{(0)}_0)+
\frac{3}{64a^4}-\frac{153}{32768a^{10}}+\cdots
\nonumber \\
F^{(1)}&=& c^{(1)}-
\frac{3}{128a^6}+\frac{657}{32768a^{12}}+\cdots
\nonumber \\
F^{(2)}&=&\frac{1}{160a^2}+
\frac{9}{1024a^8}-\frac{16749}{262144a^{14}}+\cdots \nonumber
\\F^{(3)}&=& \frac{5}{2688a^4}-\frac{3}{1024a^{10}}+\frac{96453}{524288a^{16}}+\cdots
\end{eqnarray}
For $N_f=2$ we have
\begin{eqnarray} \label{NekrasovmasslessNf2}
F^{(0)}&=& 2a^2\textrm{log}(a)+(c^{(0)}_2a^2+c^{(0)}_1a+c^{(0)}_0)
-\frac{1}{2}-\frac{1}{64a^2}-\frac{5}{32768a^{6}}+\cdots
\nonumber \\
F^{(1)}&=& -\frac{1}{6}\textrm{log}(a)+c^{(1)}
+\frac{1}{64a^4}+\frac{23}{16384a^{8}}+\cdots
\nonumber \\
F^{(2)}&=& \frac{7}{480a^2}
-\frac{7}{1024a^6}-\frac{1425}{262144a^{10}}+\cdots \nonumber
\\F^{(3)}&=& \frac{31}{8064a^4}+\frac{5}{2048a^8}+\frac{8843}{524288a^{12}}+\cdots
\end{eqnarray}
For $N_f=3$ we have
\begin{eqnarray} \label{NekrasovmasslessNf3}
F^{(0)}&=&
a^2\textrm{log}(a)+(c^{(0)}_2a^2+c^{(0)}_1a+c^{(0)}_0)-\frac{1}{64}-\frac{1}{32768a^{2}}+\cdots
\nonumber \\
F^{(1)}&=&
-\frac{1}{3}\textrm{log}(a)+c^{(1)}-\frac{1}{128a^2}+\frac{3}{32768a^{4}}+\cdots
\nonumber \\
F^{(2)}&=& \frac{11}{480a^2}+
\frac{5}{1024a^4}-\frac{109}{262144a^{6}}+\cdots \nonumber
\\F^{(3)}&=& \frac{47}{8064a^4}-\frac{1}{512a^6}+\frac{769}{524288a^{8}}+\cdots
\end{eqnarray}

\end{document}